# Imaging neurotransmitter transport in live cells with stimulated Raman scattering microscopy

**Authors**
Bryce Manifold[1†], Gabriel F. Dorlhiac[2†], Markita Landry[2,3,4,5,6*], Aaron Streets[1,5,6,7*]

**Affiliations and contact information**
[1] Department of Bioengineering, University of California, Berkeley, Berkeley, CA, 94720, USA
[2] Biophysics Graduate Group, University of California Berkeley, Berkeley, CA, 94720, USA
[3] Department of Chemical and Biomolecular Engineering, University of California, Berkeley, Berkeley, CA, 94720, USA
[4] Innovative Genomics Institute, University of California Berkeley, Berkeley, CA, 94720, USA
[5] California Institute for Quantitative Biosciences, QB3, University of California, Berkeley, Berkeley, CA, 94720, USA
[6] Chan-Zuckerberg Biohub, San Francisco, CA, 94158, USA
[7] UC Berkeley-UC San Francisco Graduate Program in Bioengineering, University of California, Berkeley, Berkeley, CA, 94720, USA

†These authors contributed equally
*Correspondence: landry@berkeley.edu; astreets@berkeley.edu

**Summary**

Chemical neurotransmission is central to neurotypical brain function but also implicated in a variety of psychiatric and neurodegenerative diseases. The release dynamics of neurotransmitters is correlated with but distinct from neuronal electrical signal propagation. It is therefore necessary to track neurotransmitter modulation separately from neuron electrical activity. Here, we present a new approach for imaging deuterated neurotransmitter molecules in the cell-silent window with stimulated Raman scattering (SRS) microscopy. Using SRS microscopy, we perform direct imaging of deuterated dopamine and GABA in PC12 chromaffin cells, and in primary hippocampal neurons, respectively, based on the carbon-deuterium vibrational frequencies. We demonstrate that SRS imaging of these isotopologues directly visualizes intracellular neurotransmitters without changing the neurotransmitters' chemical identity, and requiring custom synthesis or genetic encoding protocols. We further show that stimulation of neurotransmitter release results in an overall 20-50% intracellular neurotransmitter signal reduction in agreement with comparable neurotransmission dynamics studies, with the ability to observe inter- and intracellular variation in vesicular neurotransmitter release. Taken together, our data suggest that neurotransmitter isotopologues can serve as a commercially-available, biocompatible, and generalizable method to image neurotransmitters with deuterated molecules that are virtually chemically identical to their native counterparts.

**Keywords**
Stimulated Raman, neurotransmitters, isotopes, microscopy, neuromodulation

## Introduction

Information flow in the nervous system is mediated by both electrical and chemical transmission processes. Understanding both of these processes is necessary to study neurotypical activity and any deviations which can result in neuropathology. Electrical transmission governs intracellular electrical signal propagation, in the form of action potentials, on the millisecond timescale. Interneuronal communication, however, is mediated by small molecule (*i.e.* neurotransmitter) release at a neuronal synapse. Canonically, neurotransmitters are typically confined to the sub-micron sized synapse after action-potential-stimulated release, leading to rapid inter-neuron communication. Other neurotransmitters, known as neuromodulators, are thought to exhibit additional extrasynaptic diffusive action, which may modulate broader neuronal networks on the orders of centimeters and seconds. While studies examining action potential propagations have led to valuable insights regarding neuron signaling, the often stochastic and decoupled release of neurotransmitters at the synaptic terminals in response to such potentials makes electrical measurements a potentially tenuous proxy for understanding neurochemical mechanisms at the molecular level (Grienberger and Konnerth, 2012; Südhof, 2004). Furthermore, abnormal regulation of neuromodulator activity is associated with a wide variety of neuropathologies, including schizophrenia (Kesby *et al.*, 2018), epilepsy (Starr, 1996), Parkinson's (Warren *et al.*, 2017), Huntington's disease (Chen *et al.*, 2013), and addiction (Nutt *et al.*, 2015), among others. The study of chemical neurotransmission is, therefore, of fundamental importance to our understanding of neuropathologies.

Direct study of neurotransmission at the molecular level requires a technique of high spatial and temporal resolution. To this end, microscopy based approaches have been paramount to the study of this process. Fluorescence microscopy, specifically, has had particular salience to studying neurotransmission considering its amenability to whole organ to organism imaging, high spatiotemporal resolution, high sensitivity, and ready availability of probes and fluorophore-encoded mutants. However, fluorescence-based approaches face a major obstacle when imaging neurochemical communication, in that neurotransmitters are small molecules that do not intrinsically fluoresce in the visible wavelength range. Therefore, several approaches have been developed to enable fluorescence-based measurements of neurotransmitter activity, including synthesis of fluorescent chemical analogues, the use of genetically encoded receptors co-expressed with fluorescent proteins with affinity for the neurotransmitter of interest, and readouts of neuron activity by fluorescent reporters as a proxy for neurotransmitter release. For instance, monitoring of vesicular fusion can be accomplished through introduction of exogenous amphipathic "FM dyes" (Betz *et al.*, 1992), or by using the acidic vesicular environment to temporarily quench genetically encoded pHluorin sensors (Sankaranarayanan *et al.*, 2000) where vesicular release into the synapse neutralizes pH to allow fluorescence to occur during a vesicular fusion event. Yet, notably, neither technique reports directly on the neurotransmitter populations themselves inside the vesicles, nor whether neurotransmitters are actually released. Other genetically encoded fluorescent receptors report binding of the neurotransmitter. For example, cell-based neurotransmitter fluorescent engineered reporters (CNiFERS), linking neurotransmitter-binding G-protein coupled receptors to fluorescent readout, have been used to study a number of different neurotransmitters (Arroyo *et al.*, 2016; Lacin *et al.*, 2016). This technique, however, requires the transplantation of reporter cells into neuronal tissue and exhibits low spatiotemporal resolution relative to neurochemical signaling. To increase spatiotemporal sensitivity of neurotransmitter imaging, $GRAB_{DA}$ was recently developed by engineering dopamine receptors with circularly permuted eGFP, to be expressed in cell membranes, and has been used to image exocytosed volume transmission of dopamine (Sun *et al.*, 2018). Patriarchi *et al* concurrently produced a suite of genetic encoded sensors, operating on similar principles, called dLight (Patriarchi *et al.*, 2018). Both probes require extensive protein engineering, which is not easily translated to other neurotransmitters or non-model organisms. Moreover, these probes, to-date, lack single release-site sensitivity and can only measure neurotransmitters post-exocytosis. Another class of sensors based on near-infrared fluorescent nanoparticles (Kruss *et al.*, 2017) can be localized to the extracellular space in the brain and produce fluorescence changes upon neurotransmitter binding, allowing imaging of synaptic-scale neurotransmitter volume transmission in acute brain slices (Beyene *et al.*, 2019). These sensors have recently been used to image dopamine release from somatodendritic processes and to characterize dopamine release from individual synapses (Bulumulla *et al.*, 2022). All aforementioned fluorescence-based probes face similar inherent design challenges, requiring novel approaches to their molecular design, development, and deployment for each new neurotransmitter target. Furthermore, their use is limited to the measurement of extracellular analytes, making it difficult to study neurotransmitters in the intracellular environment.

To address the challenge of intracellular neurotransmitter imaging, direct chemical modification of neurotransmitters has been used to produce visible fluorescence, allowing tracking of the molecule itself. False neurotransmitter analogs have been under development for many years as artificial alternatives to their endogenous counterparts. These so-called fluorescent false neurotransmitters (FFN) enable intracellular tracking and subsequent release of the neuromodulators dopamine (Pereira *et al.*, 2016; Rodriguez *et al.*, 2013) and serotonin (Henke *et al.*, 2018). However, relatively large chemical changes to the native neurotransmitter molecules are necessary to produce a visibly fluorescent analog, requiring extensive validation and complex syntheses. Furthermore, it remains unclear to what extent the chemical moiety and structure modifications present in FFNs affect their biodistribution and function. Several investigations have therefore used the natural autofluorescence of certain neurotransmitters, such as dopamine and serotonin (Balaji *et al.*, 2005; Maity and Maiti, 2018; Sarkar *et al.*, 2014). While this approach provides a direct readout of the compound, the use of UV autofluorescence limits penetration depth and competes with abundant background autofluorescence of other endogenous molecules when considering *in situ* experiments, and precludes application to other non-intrinsically autofluorescent neurotransmitters.

Fluorescence, however, is not the only physical property that can be leveraged for chemical specificity. Molecular vibrations, for example, are specific to the types of atoms and bonds in a molecule and can be probed through several different optical interactions, many of which are well suited to implementation as imaging techniques. The most common techniques for probing molecular vibrations are infrared (IR) and Raman spectroscopy, although many others exist including surface-enhanced Raman scattering (SERS), photoacoustic imaging, and nonlinear techniques such as coherent anti-Stokes Raman scattering (CARS), and stimulated Raman scattering (SRS). However, vibrational imaging faces its own difficulty in isolating the vibrational signature of a single molecule of interest *in situ* in a biological system. A typical vibrational spectrum of a cell contains a substantial amount of information about cellular composition but, even with high spectral resolution, is often too complex to deconvolve into individual molecular contributions. To date, only a single native neurotransmitter, acetylcholine, has been imaged using vibrational microscopy in the frog neuromuscular junction (Fu *et al.*, 2017). The lack of subsequent development of vibrational probes for neuroscience illustrates the challenge in specifically imaging a single type of molecule in a complex cellular context.

Nonetheless, in the typical biological systems of interest, a “gap” of low endogenous vibrational signal appears in the spectrum from ~1800 - 2800 $cm^{-1}$, often called the “cell-silent window”. This range of the spectrum corresponds to the frequencies of alkyne and nitrile vibrations, functional groups that are largely absent from native biomolecules in animals. Alkyne labeling has therefore been used to study a number of processes including imaging of newly synthesized DNA by incorporation of 5-ethynyl-deoxyuridine (EdU) (Wei *et al.*, 2014; Yamakoshi *et al.*, 2011), to study mitochondria (Yamakoshi *et al.*, 2015), membrane proteins (Hu *et al.*, 2018), and drug uptake (Fu *et al.*, 2014; Wei *et al.*, 2014; Zhao *et al.*, 2017, Wong *et al.*, 2024). Such a modification was recently proposed to study dopamine; however, imaging based on the alkyne vibration has not yet been demonstrated (Nuriya *et al.*, 2021). An alternative strategy to the incorporation of alkyne functional groups is to use isotope labeling. Like in classical “spring-and-ball” systems, the frequency of a molecular vibration is inversely proportional to the reduced mass of the atoms involved. Increasing the mass of the involved atoms will therefore reduce the vibrational frequency. For example, consider two common isotopic substitutions used in mass spectrometry and nuclear magnetic resonance (NMR): $^{1}H$ to $^{2}H$ (D) or $^{12}C$ to $^{13}C$. These substitutions move the C-H vibrational modes that are normally at 2800-3000 $cm^{-1}$ into the cell-silent window. Isotope labeled compounds have been used in conjunction with vibrational imaging to study lipid and cholesterol dynamics (Alfonso-García *et al.*; 2015), proteome degradation, and Huntington protein aggregation (Shen *et al.*, 2014, Miao *et al.*, 2020), in addition to being used as a general readout of metabolic activity (Shi *et al.*, 2018). Furthermore, due to their use in mass spectrometry and NMR as analytical reference standards, many neurotransmitter isotopologues are already commercially available. This commercial availability provides a readily accessible source of neurotransmitters with vibrational signatures that can be easily separated from the cellular background, obviating the need for special syntheses, or genetic engineering for their implementation in neuroscience research. We hypothesize that when introduced into neurobiological systems or cell cultures, these compounds are internalized by endogenous neurotransmitter transporters (Figure 1a) given the relatively minor chemical change of the isotopologue. Indeed, deuterated dopamine, for example, has been shown to be taken up by retinal neurons and prevent myopia equivalently to nondeuterated dopamine (Thompson *et al.*, 2020). Perhaps more convincingly, deuterated neurotransmitter isotopologues have been shown to have increased anti-akinetic potency without increased dyskinesias in rodents (Mamlöf *et al.,* 2008; Mamlöf *et al.,* 2010; Mamlöf *et al.,* 2015) and longer biological half-life in both rodents and humans (Schneider *et al*., 2018) for the treatment of Parkinson's disease with Phase I trials reportedly complete (Teva Pharmaceuticals International GmbH, 2024). Taken together, previous work clearly shows deuterated isotopologues are recognized by cellular machinery and functional *in vivo*. For our imaging purposes here, vesicular populations of the internalized deuterated neurotransmitter of interest can be monitored under native conditions using vibrational microscopy.

Specifically, we will demonstrate imaging of these compounds using SRS microscopy. SRS is a coherent third order nonlinear process that makes use of two pulsed lasers and is implemented in a point-scanning configuration. In SRS, the two laser frequencies (i.e. wavelengths) are chosen such that the difference in their frequency is resonant with a vibrational mode of interest. Through coherent driving of the vibration, signal enhancements of ~$10^8$ over spontaneous Raman scattering can be obtained (Freudiger *et al*., 2008; McCamant *et al*., 2003, Gao *et al*., 2023).

In this report, we demonstrate the application of SRS microscopy with neurotransmitter deuterium labeling to image two such commercially-available deuterated neurotransmitters, dopamine-$d_4$ (DA-$d_4$) and γ-aminobutyric acid-$d_6$ (GABA-$d_6$), in cultured PC12 cells and primary hippocampal neurons, respectively. We show that leveraging the signal enhancement of the SRS process enables visualization of DA-$d_4$ and GABA-$d_6$ neurotransmitter isotopologues at speeds comparable to confocal fluorescence microscopy, with molecular structures that are functionally identical to their native counterparts.

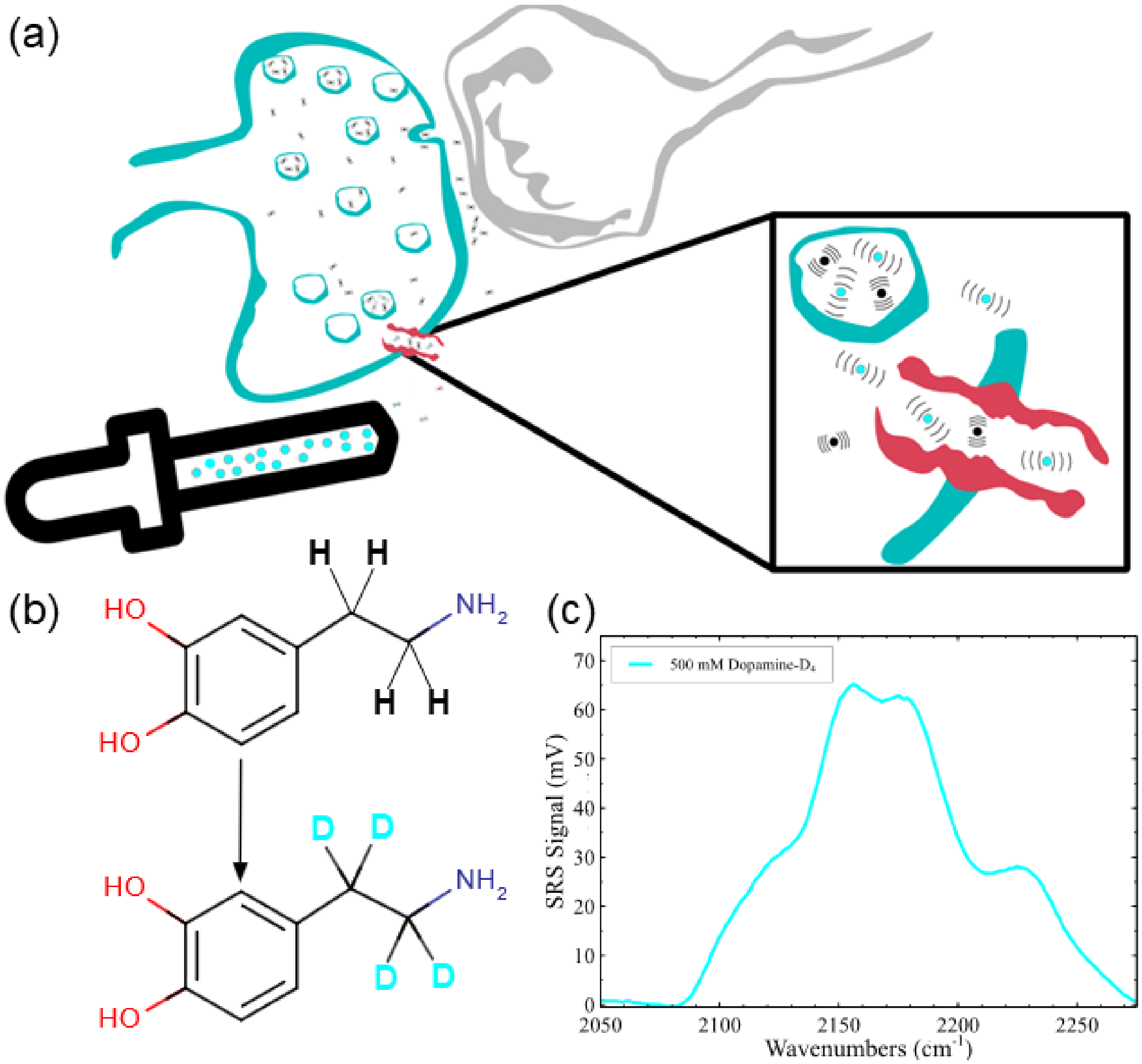

**Figure 1. Neurotransmitter Isotopologue imaging with SRS microscopy and DA-d4.**

Panel (a) shows a synapse where extracellular deuterated neurotransmitter isotopologues (cyan dots, lower frequency vibrations) can enter the cells and be packaged in vesicles like native neurotransmitters (black dots, higher frequency vibrations). Panel (b) shows a native dopamine molecule (top) and the deuterated isotopologue version dopamine-$d_4$ (bottom) used here. Panel (c) shows the stimulated Raman scattering spectra in the CD region of dopamine-d4.

## Results

<u>SRS microscopy facilitates observation of DA-$d_4$ uptake and release</u>

As a first demonstration, we investigated the capability to image the uptake and intracellular distribution of the deuterated dopamine neurotransmitter, DA-$d_4$, in cultured

PC12 cells. The PC12 cell line, derived from rat chromaffin cells, is a popular *in vitro* culture system for studying catecholamines and neurite development (Westerink and Ewing, 2008). Figure 1b shows the hydrogen-to-deuterium substitutions made for this particular isotopologue, while its corresponding SRS spectrum acquired in the cell-silent window is displayed in Figure 1c. We first tested the ability to detect uptake of the labeled compound DA-$d_4$, by adding it to a final concentration of 50 µM to the cell culture medium, notably far below the sensitivity limit of ~2mM for detection in SRS (Figure S1). The PC12 cells were incubated for one hour, after which they were washed and then imaged in PBS.

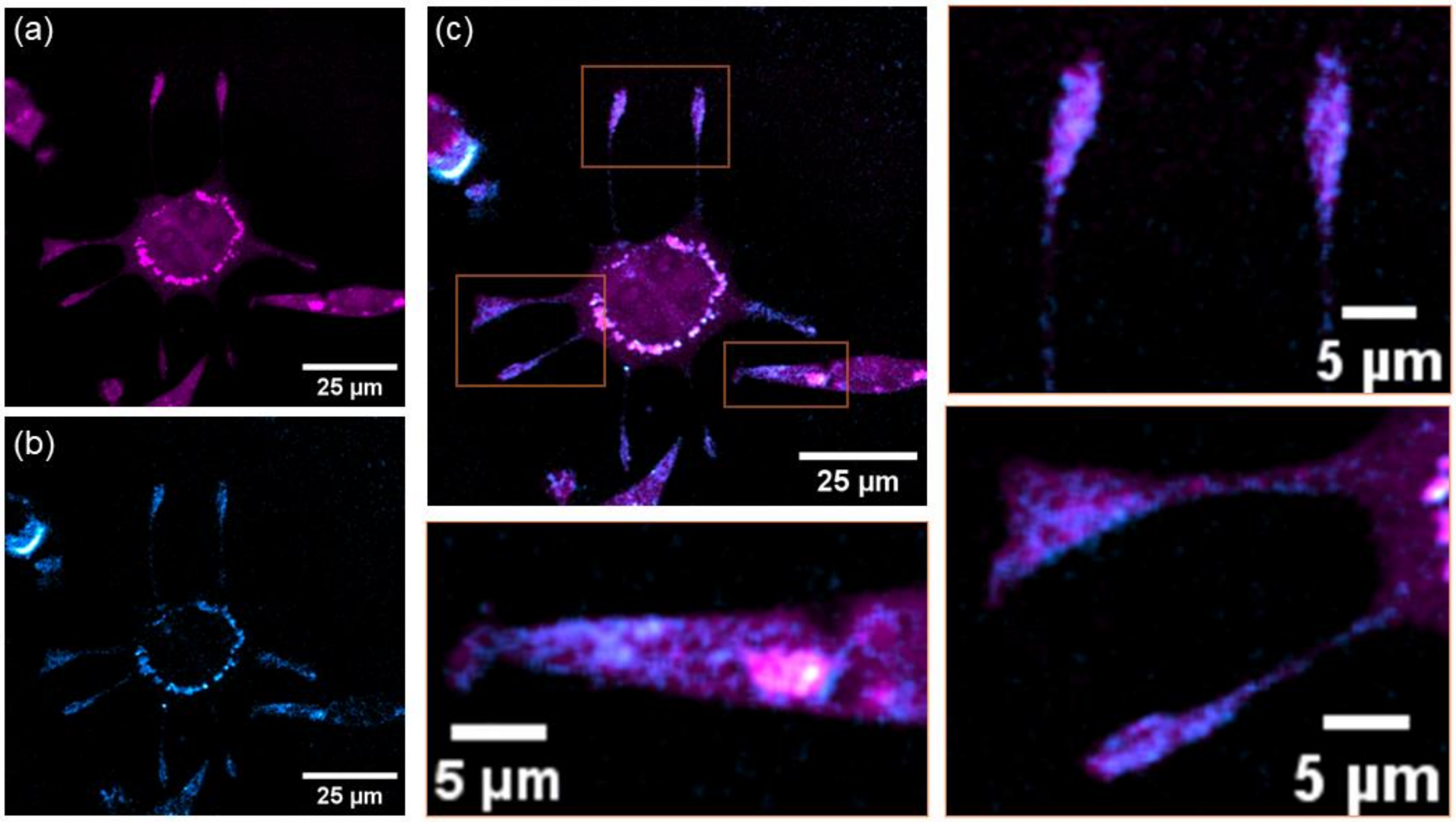

**Figure 2. SRS images of PC12 Cells dosed with 50 µM DA-d4.**
Panel (a) shows a CH SRS image (~2930 $cm^{-1}$, magenta) of live PC12 cells that have been dosed with DA-d4. Here the SRS signal comes from endogenous biomolecules with carbon-hydrogen bonds, showing typical features of the cells and their organelles. Panel (b) shows the same field of view in the CD region (~2150 $cm^{-1}$, cyan). Here the SRS signal only comes from the dosed DA-d4 accumulation within the cell. Panel (c) and insets show the CH and CD images composited together to show the localized accumulation of DA-d4 within the cells, especially in the neurite and axon like outgrowths. Close inspection reveals point-like features in the CD images which are likely the vesicles in which the deuterated dopamine is stored at sufficiently high concentration for detection in SRS.

To verify uptake, images were acquired at 2150 $cm^{-1}$, the maximum of the DA-$d_4$ SRS spectrum in the cell-silent window (Figure 1c), and compared to the images of cells which were not incubated with DA-$d_4$. To provide a reference for the location and structure of the entire cell, images were also acquired at 2930 $cm^{-1}$, corresponding to symmetric $CH_3$ stretches, a signal largely dominated by proteins. DA-$d_4$ internalization was observed for the DA-$d_4$-incubated cells, largely localized to the cell periphery, and in punctate structures (suggesting vesicular uptake of the DA-$d_4$), with clear nuclear exclusion (Figure 2). To provide further confirmation of uptake, spectra were acquired of incubated and control cells further demonstrating DA-$d_4$ uptake (Figure S2). Under identical imaging and culture conditions, unincubated cells displayed no such signal, consistent with detection of a signal specific to the DA-$d_4$ neurotransmitter (Figure S3).

The observed subcellular distribution of DA-$d_4$ reveals signal along the cell periphery and in distinct puncta as predicted, further suggesting a biologically-relevant internalization mechanism, with the isotopologue confined to intracellular vesicles. Comparatively little to no signal is observed outside these puncta, consistent with a much lower concentration of neurotransmitter present in the cytoplasm, estimated to be 0.5 – 5 μM (Jones *et al.*, 1999; Olefirowicz and Ewing, 1990), compared to 110 – 190 mM in the vesicles (Chen *et al.*, 1994; Wightman *et al.*, 1991). Furthermore, we observe the expected lack of DA-$d_4$ internalization into the nucleus. We do note here the colocalization of DA-$d_4$ signal with bright puncta in the CH images. These are potentially features of colocalized bright CH signals such as lipids, lysosomes, or organelles that may associate with the DA-$d_4$ in vesicles or otherwise. We also note that DA-$d_4$ is not entirely free of CH Raman signal with its remaining aromatic carbon-hydrogens (Figure 1B and Figure S2). To test whether this DA-$d_4$ signal arises from an active vesicular population of DA-$d_4$, we incubated PC12 cells with 50 μM DA-$d_4$ and subsequently stimulated DA-$d_4$ release with 50 mM potassium. The cells were then imaged two minutes post-stimulation. In line with the expected response to potassium stimulation, there was an observable decrease in the measured DA-$d_4$ signal for the population of cells imaged post stimulation (Figure 3a and 3b), which is significant compared to the unstimulated cells (Figure 3c; $p < 0.001$, $\Delta > 0.35$). This 22% decrease is commensurate with prior reports that quantal release of dopamine in PC12 cells is fractional (Rodriguez *et al*., 2013; Sarkar *et al*., 2014.). We also observe a significant difference between the average intracellular DA-$d_4$ signal intensity in incubated vs unincubated cells (Figure 3c; $p < 0.001$, $\Delta > 15$). Additional fields of view depicting the cells used in these population metrics are shown in Figures S3 - S5.

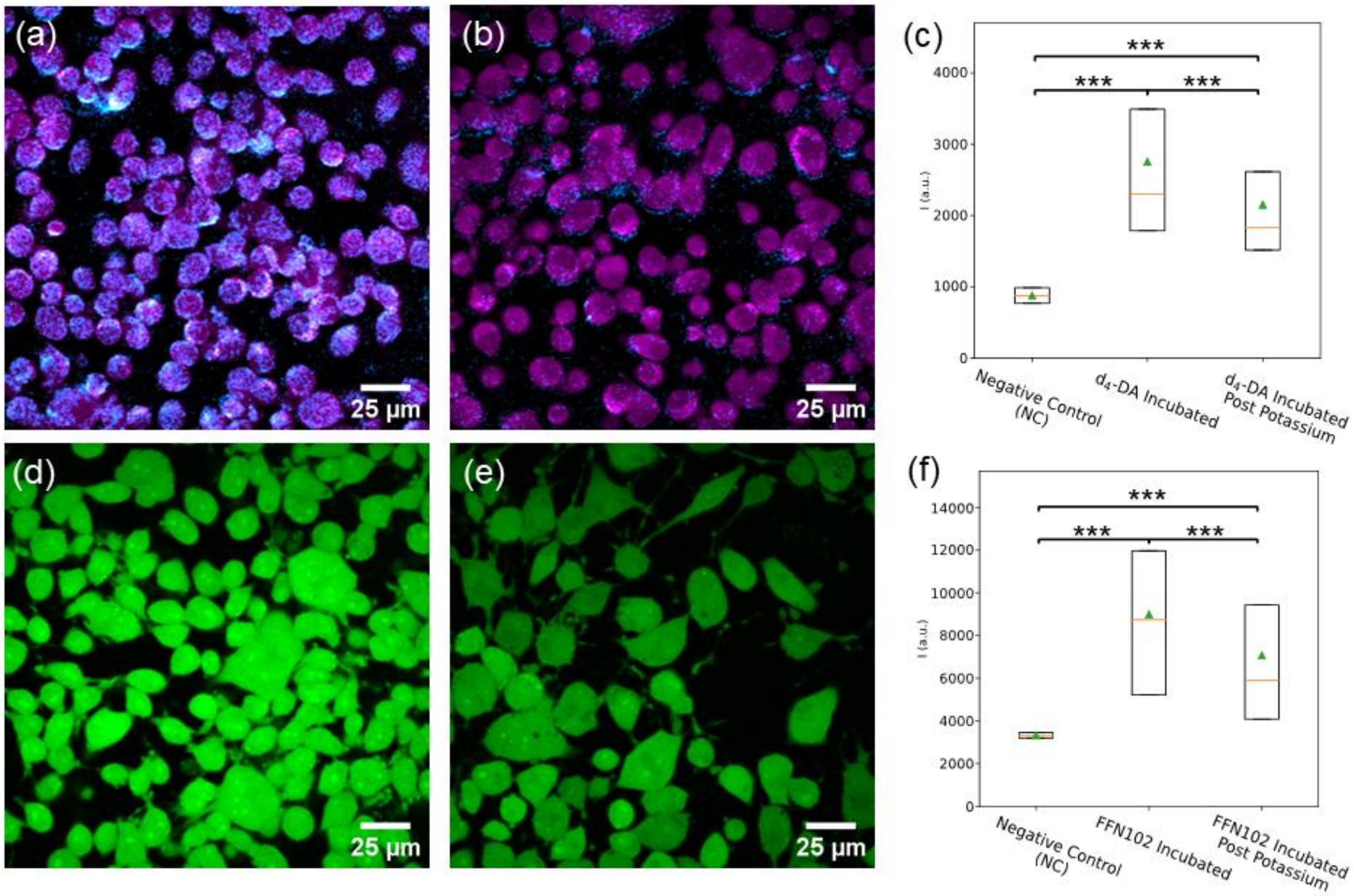

**Figure 3. PC12 uptake and release of DA-$d_4$ and FFN102 under $K^+$ stimulation**
Panels (a) and (b) show composite SRS images of PC12 cells dosed with 50 µM DA-d4 before and after stimulation with 50 mM K+, respectively. Here the CH (endogenous) signal is shown in magenta, while the CD (DA-d4) signal is shown in cyan. Panel (c) shows the population box plots of average cellular CD SRS signal for PC12 cells under negative control (n = 1273), DA-d4 incubation only (n = 1152), and DA-d4 incubation with subsequent K+ stimulation (n = 1145). Panels (d) and (e) show two-photon fluorescence images of PC12 cells dosed with FFN102, a fluorescent small molecule thought to mimic dopamine uptake and release, before and after 50 mM K+ stimulation, respectively. Panel (f) shows the population box plots for two-photon fluorescence intensity of PC12 cells under negative control (n = 1802), FFN102 incubation only (n = 1904), and FFN102 incubation with subsequent K+ stimulation (n = 2920).

We also compared the performance of a dopamine FFN, FFN102, in the same potassium stimulation experiments. Here, PC12 cells were incubated with 10 µM FFN102. A strong signal was visible within cells incubated with FFN102 when imaged by two-photon excited fluorescence (TPEF) microscopy. (Figure 3d). Notably, there was stronger FFN102 signal from the entire cell and unlike DA-$d_4$ labeled cells, FFN signal was observed diffusely in the cytoplasm and in the cell nuclei, suggesting at least some non-specific internalization mechanisms of the dopamine FFN analogue. FFN102-incubated cells imaged post-potassium stimulation also displayed a lower overall signal relative to non-potassium stimulated controls (Figure 3e). This 21% decrease for FFN-incubated cells post stimulation was also observed to be significant over a large sample of cells (Figure 3f; $p < 0.001$, $\Delta > 0.48$). This change is consistent with the 25% reduction in autofluorescence of DA (Sarkar *et al.* 2014) and our observed 22% reduction in CD signal of DA-$d_4$ under $K^+$ stimulated release. Taken all together, this suggests that SRS imaging of DA-$d_4$ can report on relevant uptake and release dynamics in PC12 cells like previously demonstrated systems.

PC12 cells display cell-to-cell heterogeneity of DA-$d_4$ uptake and release

We next sought to measure stimulated release in individual cells that were incubated with DA-d4, or FFN102. While population level measurements have relevance for several important biological questions, they often obscure cell-to-cell heterogeneity. It is, therefore, frequently of interest to take repeat measurements on the same individual cells. Single-cell measurements are of particular relevance for neurons, where measuring the effect of stimulation and the relevant kinetics of neurotransmitter release are paramount to understanding neuro-communication. Furthermore, in PC12 cells, it has been shown previously that multiple vesicular populations (Westerink *et al.*, 2000) can exist heterogeneously across individual cells, producing cell-to-cell variability in dopamine concentration that is often not apparent when making bulk population level measurements (Westerink and Ewing, 2008).

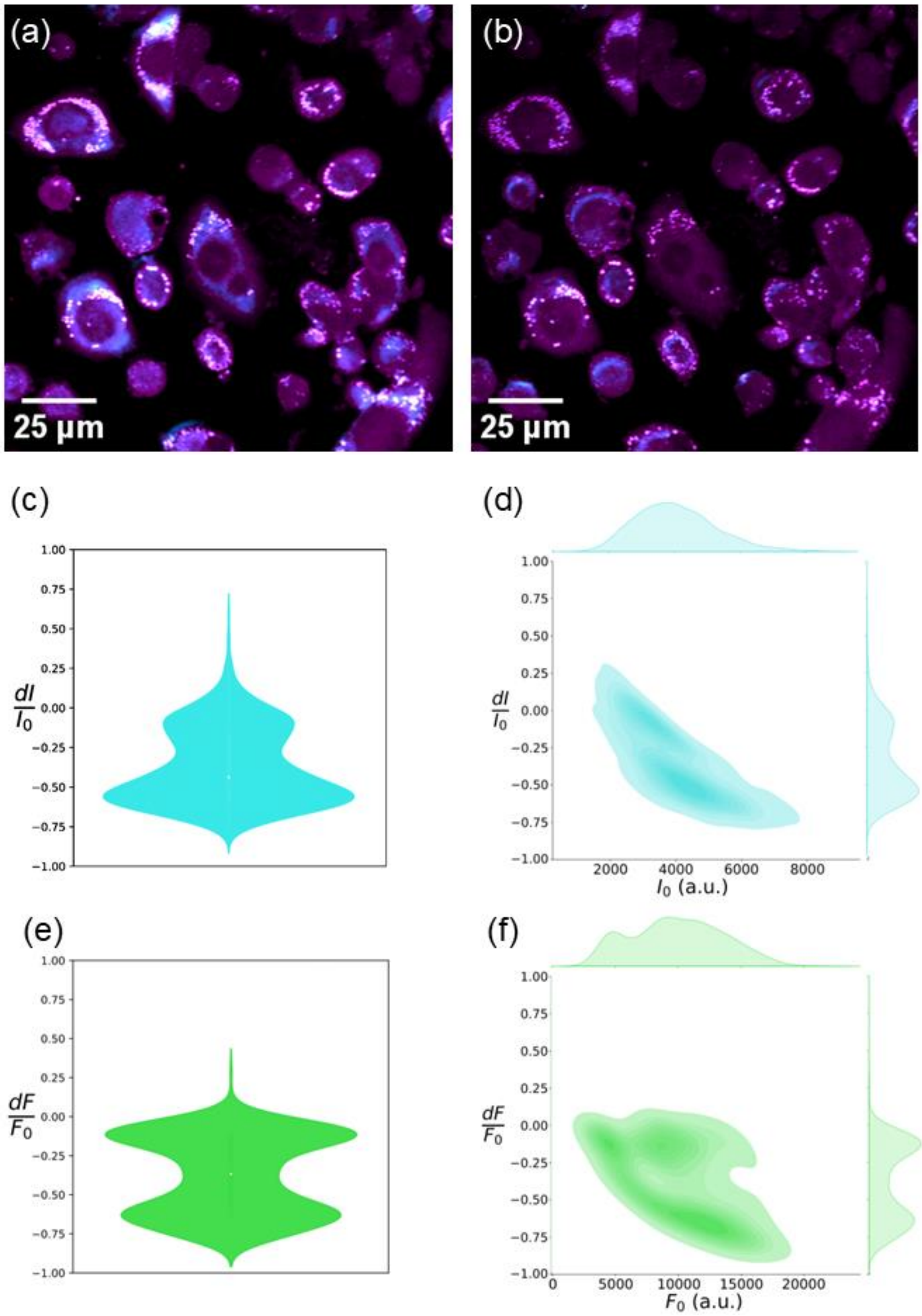

**Figure 4. Cell-to-cell comparison of DA-d4 and FFN102 uptake and release.** Panel (a) and (b) show PC12 cells dosed with DA-$d_4$ before and after stimulation with $K^+$ solution, respectively. Panel (c) shows the cell by cell comparison of change in normalized CD signal versus initial CD signal in the potassium stimulation experiment. Panel (d) shows the same as (c), but additionally plotted versus the initial intensity of the CD signal. Panels (e) and (f) are corollary to (c) and (d), respectively, but with two photon fluorescence signal instead.

To measure DA-$d_4$ concentration within the same PC12 cells pre- and post-stimulation, we built a microfluidic flow cell, with inlets for the injection of stimulating potassium solution. Cells were grown inside these devices, composed of coverslips on the top and bottom, and PDMS around the sides. Using these flow cells, we first incubated PC12 cells with 50 μM DA-$d_4$, as in prior experiments. Prior to imaging, the cells were washed and imaging was performed in fresh culture medium. As before, we first acquired images at 2930 $cm^{-1}$ to localize cell boundaries and to serve as a mask to identify and separate cells computationally during image analysis. We then acquired images of the same fields of view at 2150 $cm^{-1}$ to image intracellular DA-$d_4$ before and after stimulation by exposure to a 50 mM $K^+$ solution (Figure S6). We next quantified the relative change in DA-$d_4$ intensity, i.e. the change in DA-$d_4$ intensity normalized to the initial intensity, in response to potassium stimulation on a single-cell basis. This analysis revealed a bimodal distribution for PC12 cells incubated with DA-$d_4$, as seen in Figure 4c, with one cell population exhibiting a strong release of DA-$d_4$ in response to potassium stimulation (64% reduction in signal), and another population exhibiting a weaker release of DA-$d_4$ (15% reduction in signal). This bimodal distribution is similar to the distribution observed when the experiment is repeated for PC12 cells incubated with FFN102 (Figure 4e). We additionally examined whether a cell's initial DA-$d_4$ intensity (a proxy measurement of intracellular dopamine concentration) is correlated to its subsequent release magnitude by comparing the relative change in DA-$d_4$ intensity due to $K^+$ stimulation as a function of the initial DA-$d_4$ intensity (Figure 4d). We observe a general trend where cells showing higher DA-$d_4$ intracellular signals show greater release – i.e. higher change in DA-$d_4$ intensity in response to potassium stimulation in both distinct populations. Interestingly, while we also observe these two populations for FFN102-incubated cells, a third, high initial intensity, but weakly-releasing, cell population is also present in FFN102-incubated cells (Figure 4f). This result further points to a nonspecific internalization mechanism of the FFN102 as is also evidenced by the observed 2-photon fluorescence images. We note that such nonspecific internalizations are not observed in the DA-$d_4$ system.

GABA-$d_6$ uptake and release can be observed with SRS microscopy

We next assessed whether the strategy of neurotransmitter deuteration combined with SRS microscopy could be applied more broadly to different commercially-available deuterated neurotransmitters. GABA, for example, plays a crucial role in the nervous system as the main inhibitory neurotransmitter and is consequently centrally implicated in many neurochemical processes throughout the brain. Furthermore, while dopamine as a neuromodulator exhibits volume transmission-based activity, GABA is thought not to, representing a class of neurotransmitter putatively confined to the neuronal synapse. While recent advances have been made for imaging GABA with exogenous fluorophores in FRET-based biosensors (Lecat-Guillet *et al.*, 2017; Masharina *et al.*, 2012), and a genetically-encoded GABA biosensor (Marvin *et al.*, 2019), no tools or molecular analogues, to our knowledge, have yet been developed for direct GABA imaging. To test the capability of SRS microscopy to image the uptake and stimulated release of a deuterated GABA isotopologue we used dissociated fetal hippocampal neurons from rats.

Here the isotopologue used, GABA-$d_6$, has all of its aliphatic hydrogens replaced with deuteriums (Figure 5a). Fetal hippocampal neurons, dissociated from day 19 Sprague Dawley rat embryos, were incubated with 50 μM GABA-$d_6$ prior to imaging.

While GABA-$d_6$ has a slightly different spectral signature than DA-$d4_4$ in the cell-silent window (Figure 5b), the strongest peak at 2150 $cm^{-1}$ was again chosen for imaging. Images were again acquired in the 2930 $cm^{-1}$ channel to provide contrast of the cell body.

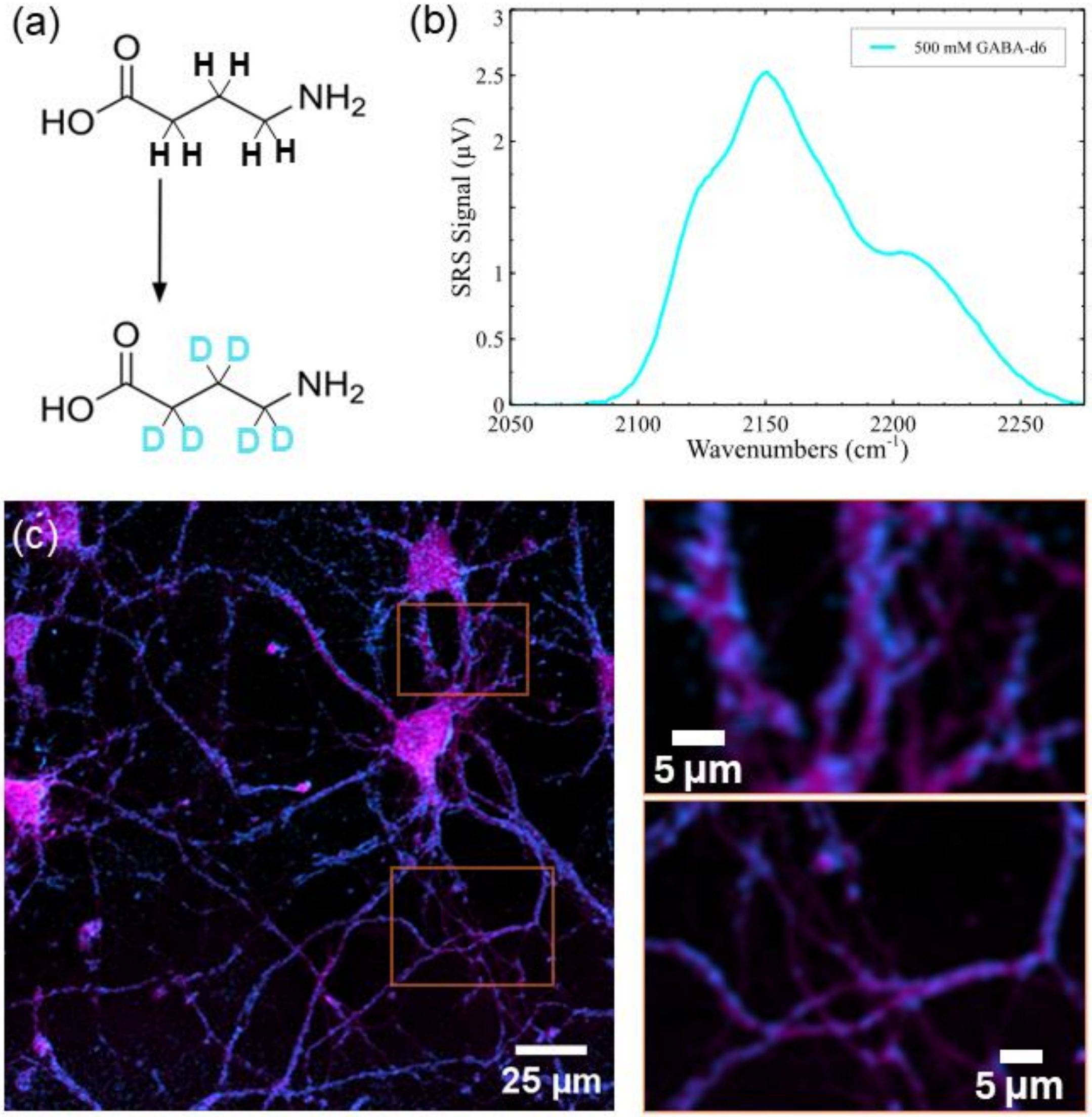

**Figure 5. Observing GABA-d6 in rat hippocampal neurons with SRS microscopy.** Panel (a) shows the native GABA molecule (top) and the deuterated GABA-d6 isotopologue. Panel (b) shows the CD SRS spectra of GABA-d6, native GABA, and phosphate buffered saline solutions. Panel (c) shows composite CH (magenta, 2930 $cm^{-1}$, endogenous biomolecules) and CD (cyan, 2150 $cm^{-1}$, intracellular accumulated GABA-d6) SRS images of live dissociated rat hippocampal neurons with insets zoomed in on the axon projections of the neurons. Here, the small cyan puncta of CD signal are again seen, evocative of intravesical GABA-d6 within these cells.

Hippocampal neurons displayed a strong signal at 2150 $cm^{-1}$ when incubated with the deuterated neurotransmitter (Figure 5c). Here the CD signal is once again observed in puncta within the cells, particularly evident along the axonal features (Figure 5c insets). It is important to note, however, that the vesicles in these cultured neurons are likely much smaller than in the PC12 cells. The size of vesicles here are likely to be tens to a few hundred nanometers in diameter, far below the point spread function of our microscope system. This results in uncertainty spatially and with respect to absolute concentration within individual vesicles. Integrating over the whole cell-body, however, does reveal significant population-level observation of internalized GABA-$d_6$. Additional fields of view are shown in Figures S7 and Figure S8. As expected, GABA-$d_6$ signal was absent from images acquired from hippocampal neurons not incubated with GABA-$d_6$ but otherwise cultured under identical conditions (Figure S9). Across many neurons, we observe a significant separation between the average signal intensity in GABA-$d_6$ incubated vs unincubated cells (Figure 6c; $p < 0.001$, $\Delta > 5.58$), confirming uptake of GABA-$d_6$ in hippocampal neurons and confirming the selectivity of SRS for imaging GABA-$d_6$ at 2150 $cm^{-1}$. In contrast to the case of PC12 cells, however, we note a population of cells do not uptake GABA-$d_6$, reflected in an overlap between the incubated and unincubated distributions. This large variation in GABA-$d_6$ uptake by hippocampal neurons is likely due to the many neuronal subtypes that exist in the hippocampus where glutamatergic neurons may be more prevalent than GABAergic neurons (Zhu *et al.*, 2017). Evidence also suggests that glutamate and GABA may coexist and may be co-released in certain neurons, which could affect relative GABA uptake (Shabel *et al.*, 2014; Zhu *et al.*, 2017). For such subtypes, a lack of or decreased uptake of GABA-$d_6$ would be expected.

Finally, to again confirm whether the GABA-$d_6$ signal measured at 2150 $cm^{-1}$ arises from active internalization, we repeated the stimulated release experiments by incubating neurons with 50 μM $d_6$-GABA for one hour then stimulating with 50 mM potassium for 2 minutes. Images were acquired in PBS after first washing away the potassium and $d_6$-GABA containing medium. Neurons imaged after potassium stimulation showed little to no observable signal at 2150 $cm^{-1}$ (Figure 6b). This 55% decrease in $d_6$-GABA signal following potassium stimulation was found to be significant across many cells (Figure 6c; $p < 0.001$, $\Delta > 0.65$). Furthermore, we compared $d_6$-GABA signal in post-potassium stimulated cells, and neurons not incubated with $d_6$-GABA. We found no significant difference in signal between $d_6$-GABA neurons imaged post potassium stimulation, and unincubated neurons, suggesting successful and near complete release of internalized $d_6$-GABA. Together these data suggest the ability to visualize uptake and release of a GABA isotopologue within cultured primary neurons. To the best of our knowledge, these results represent the first optical micrographs of the neurotransmitter isotopologue of GABA within a neurobiological system.

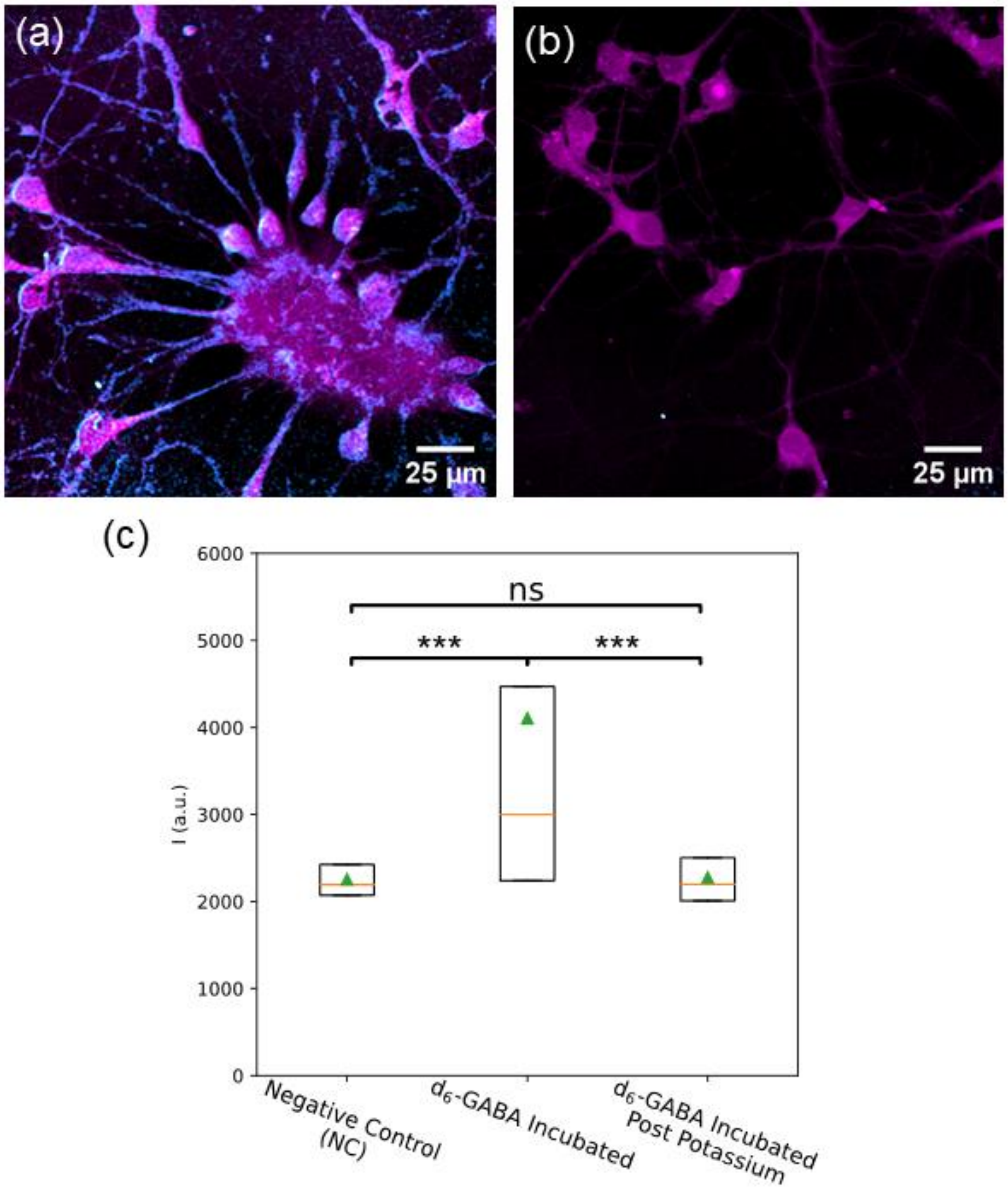

**Figure 6. Primary hippocampal neurons uptake and release GABA-d6.** Panel (a) and (b) show CD SRS images of live rat hippocampal neurons dosed with 50 µM GABA-d6 before and after stimulation with 50 mM K+ solution, respectively. Panel (c) shows population box plots of average cellular CD SRS signal under the conditions of negative control (n = 1840), GABA-d6 incubation only (n = 2110), and GABA-d6 incubation with subsequent K+ stimulation (n = 2537).

## Discussion

Isotopes have a long history of use in neuroscience. Some of the first techniques developed to measure concentrations of neurotransmitters were based on radioactive isotopes (Bowdler *et al.*, 1983; Lieberman *et al.*, 1969). Positron emission tomography (PET) continues to make frequent use of radiolabels to study the proteins of the nervous system (Chen *et al.*, 2020; Stehouwer and Goodman, 2009). Powerful mass spectrometry imaging methods use deuterated analogues as calibration standards for quantitative spatial mapping of neurotransmitters (Fernandes *et al.*, 2016; Shariatgorji *et al.*, 2014; Zhu *et al.*, 2011). In this report, we extend the use of isotopes in neuroscience and demonstrate that by combining the use of commercially-available isotopologues with vibrational microscopy. We can image the biodistribution and stimulated release of deuterated dopamine and GABA isotopologues in cultured cells and primary neurons with sub-cellular resolution. Our results suggest that commercially-available deuterated neurotransmitters and neuromodulators could be applied to a broad range of studies of neurotransmitter uptake and release kinetics in neuronal cells, tissue, and potentially in living animals. Recent developments in the field of vibrational microscopy also provide opportunities to further enrich the information this technique provides through integration of orthogonal methodologies that simultaneously measure neurotransmitter release kinetics and electrical activity. Further, the higher spectral specificity of vibrational microscopy, in comparison to fluorescence microscopy, allows for easier uptake and release observation of multiple neurotransmitters simultaneously.

SRS microscopy, for instance, has been shown to be sensitive to membrane potential (Liu *et al.*, 2015), a property recently exploited to track neuron depolarization in response to action potentials (Lee *et al.*, 2017). In combination with isotopologue imaging, SRS offers the potential for a single optical system to image both neurochemical and electrical information simultaneously. The narrow linewidths of Raman transitions (~10 $cm^{-1}$), compared to fluorescence emission spectra (~$10^5$ $cm^{-1}$), raise the additional possibility of simultaneous observation of multiple neurotransmitters by tuning vibrational peaks with various isotope labelings with no additional optical setup complexity (Wei *et al*. 2017). Acquisition of hyperspectral image stacks, i.e. multiple images of the same field of view at many Raman transitions, would allow for the spectral unmixing of the contribution of distinct neurotransmitter isotopologues. A multiplexed imaging strategy could help answer questions surrounding release and uptake of multiple neurotransmitters in single neurons, as in the case of GABA and glutamate, two neurotransmitters that canonically have opposing activity (Shabel *et al.*, 2014; Zhu *et al.*, 2017).

The current study has demonstrated isotopologue imaging in cells and primary neurons with transmission-based SRS imaging. To extend the use of neurotransmitter isotopologues to tissue samples, or even live animals, an epi-SRS microscope could be used, with a geometry analogous to more familiar fluorescence-based microscopes. The use of epi-SRS has previously been demonstrated to work in brain tissue both *ex vivo* (Bae *et al.*, 2018; Ji *et al.*, 2013; Manifold *et al.*, 2019) and *in vivo* (Francis *et al.*, 2022). In fact, SRS imaging is rather unique in this regard, as signal collection in the epi-direction increases with increasing sample thickness. For samples more than 2 mm thick, it has been found to be preferable to image in this backscattering geometry (Hill & Manifold *et al.*, 2020). Additionally, other vibrational imaging techniques are not subject to the same constraints. For instance, CARS, another coherent Raman technique, is frequently implemented in either a transmission or epi geometry, and confocal spontaneous Raman and SERS microscopy are almost exclusively implemented in epi geometries. As such, there are numerous opportunities to adopt existing Raman-based imaging techniques for isotopologue imaging across a broad range of neurobiological studies of neurotransmission.

Similarly, exclusively intracellular populations of dopamine and GABA were measured using our current imaging setup, since neither spontaneous Raman nor SRS microscopy are likely to achieve the sensitivity required to observe the lower concentrations of neurotransmitters present extracellularly. For neuromodulators like dopamine, for which the study of volume transmission is of interest, their extracellular concentrations can vary widely but are often present at < 20 nM concentrations extracellularly (Watson *et al.*, 2006). Isotopologue imaging for intracellular studies combined with existing probes to image volume transmission would present a powerful opportunity to track neurotransmitter uptake, release, and volume transmission. As such, the current combination of neurotransmitter isotopologues and SRS microscopy is best implemented to study intracellular neurotransmitter populations and can serve as an orthogonal measure, complementing other tools available to study neuromodulator volume transmission. These include direct extracellular concentration sensors, such as those based on carbon nanotubes or genetically engineered proteins, or techniques like magnetic resonance spectroscopic imaging (MRSI), which is used to study neurotransmitter concentrations under a variety of paradigms. While powerful, MRSI is unable to distinguish intracellular and extracellular contributions to the signal (Spurny *et al.*, 2020), due to lower spatiotemporal resolution, a void which our technique may help to fill when questions of mechanism or attribution arise. Additionally, signal enhancement techniques may also be used to further increase the sensitivity of vibrational microscopy. Recent work suggests that surface enhanced SRS (SESRS) can be used to image at the single molecule level as demonstrated with bacterial cells (Zong *et al.*, 2019). Similar observations have been made for surface enhanced CARS (SECARS) (Koo *et al.*, 2005; Steuwe *et al.*, 2011; Zhang *et al.*, 2014). Furthermore, a recent study (Zong *et al.*, 2021) describes a widefield implementation of SECARS which could extend that level of sensitivity to the millisecond temporal resolution of more conventional widefield microscopies. While the use of plasmonic nanoparticles, or culturing on plasmonic nanostructure surfaces, introduces an additional level of complexity, the SERS field has frequently demonstrated its applicability to and utility in neuroimaging (Diaz *et al.*, 2014; Nicolson *et al.*, 2019), even in intraoperative/clinical settings (Jiang *et al.*, 2019; Karabeber *et al.*, 2014).

In this report we have presented a practical strategy to image neurotransmitter isotopologues under native biological conditions using vibrational microscopy. This strategy generalizes to any class of neurotransmitter, many of which are already commercially available in isotope-labeled form, opening new avenues for fundamental investigations into the biology of neurotransmitters. The existing and long-standing use of deuterated neurochemicals in clinical practice further substantiates the biocompatibility of their use and motivates the use of our approach in non-model organisms and human brain tissues. Notably, many commercially-available isotopologues, such as GABA as demonstrated here, are neurotransmitters or neuromodulators that currently lack cellular-scale imaging tools. While this work represents the first demonstration of the use of isotopologues and vibrational microscopy in this context, it makes use of a technique, *i.e.*, isotope labeling, which is broadly familiar to both the microscopy and neuroscience communities. Furthermore, a growing demand for new tools to study the nervous system has led to increasing adoption of vibrational imaging methods in neuroscience, which can be easily extended to the study of neurotransmitter uptake and release as demonstrated herein. Taken together, our results suggest isotope labeling of neurotransmitters, combined with SRS microscopy, represents a strategy with the potential for wide adoption to further the understanding of neurotransmission in the brain.

**Acknowledgments and funding**

We would like to acknowledge Professor Evan Miller and his students Pavel Klier and Deshka Lynn Niell for providing primary hippocampal cells, and advice on maintenance of their culture conditions. Co-corresponding authors contributed equally to this study and may modify ordering of author positions in the author list when presenting this work. We acknowledge support of a Burroughs Wellcome Fund Career Award at the Scientific Interface (CASI) (MPL), a Dreyfus foundation award (MPL), the Philomathia foundation (MPL), an NIH MIRA award R35GM128922 (MPL), a NIH R21 NIDA award 1R03DA052810 (MPL), a NSF CAREER award 2046159 (MPL), a NSF CBET award 1733575 (to MPL), a CZI imaging award (MPL), a Sloan Foundation Award (MPL), aUSDA BBT EAGER award (MPL), a Moore Foundation Award (MPL)., a National Science Foundation grant 1845623 (AS). AS is a Chan Zuckerberg Biohub Investigator and a Pew Biomedical Scholar. MPL is a Chan Zuckerberg Biohub investigator, a Hellen Wills Neuroscience Institute Investigator, and an IGI Investigator.

**Author contributions:**

Conceptualization: AS, MPL
Methodology: GFD, BM, AS, MPL
Investigation: GFD, BM
Visualization: BM, GFD
Supervision: AS, MPL
Writing—original draft: GFD
Writing—review & editing: BM, GFD, AS, MPL

**Competing interests:** The authors declare that they have no competing interests.

**Data and materials availability:**

Image data are presented in the manuscript and supplement. Additional fields of view available upon request by emailing the corresponding authors.

**Materials and Methods**

<u>PC12 cell culture</u>

PC12 rat chromaffin cells (UC Berkeley cell culture facility) were continuously cultured in RPMI-1640 medium with L-glutamine, containing 10% goat serum, 10% FBS, and 1% penicillin/streptomycin. For DA-$d_4$ imaging, cells were passaged onto 25 mm diameter round coverslips. For FFN102 imaging, cells were passaged into 24 well-plates. Upon achieving 50-70% confluency on the coverslips, the media was exchanged to a serum deprived DMEM media, containing 0.5% FBS, 1% penicillin/streptomycin, and 100 ng/mL NGF-β. The coverslips were imaged 5 days post media exchange. The cells were maintained in an incubator at 37 °C and 5% $CO_2$.

Flow cell construction and PC12 cell culture on device

SRS imaging of DA-$d_4$ in cells introduces a set of strict geometric constraints for sample placement on the microscope. This is due to the fact that the SRS signal is, generally, collected in a transmission geometry (Figure S10), requiring the sample be placed between short working-distance objective and condenser lenses. SRS microscopy can also be subject to non-specific signals arising from the optical Kerr effect in the form of cross-phase modulation (XPM). The most straight-forward way to minimize this effect is to use a collecting lens with a numerical aperture equal to or larger than that of the objective lens (Berto *et al.*, 2014), necessitating a 1.4 NA oil immersion condenser lens, for use with the 1.2 NA water immersion objective used in this study. In order to make measurements of DA-$d_4$ signal intensity on the same PC12 cells both before and after potassium stimulation, we constructed a simple flow cell.

The flow cell was constructed using two 48 x 60 mm coverslips (Gold Seal, Thermo scientific). RTV 615 PDMS was first mixed in a 10:1 ratio of monomer to cross-linker. PEEK tubing (1/32”x0.010, IDEX Health and Science) was cut into lengths of 2” to serve as inlets into the device, and to control the height of the device. Two lengths of tubing were placed in the center of one short edge of the coverslip, while another two were placed along the entire length of each long edge. The PDMS mixture was then applied using a brush along the two long edges and the short edge with tube inlets. The PDMS was applied in a thickness equal to the height of the PEEK tubing extending up from the coverslip. The second coverslip was then placed on top, enclosing the device. The device was baked for 10 minutes at 70 °C, then flushed with ethanol and further sterilized under UV-C light for 2 minutes. The final height of the devices was ~1.1 mm.

PC12 cells were introduced to the device by extracting ~2 mL of trypsanized cells from the passaging culture using a syringe, sterilized as above using ethanol and UV-C light, and then affixing the syringe to one of the available inlets. Cells were allowed to grow for 2 days before the medium was exchanged for the serum deprived medium. Cells were grown for a further 5 days before imaging, the same as those on coverslips. The devices were maintained within separate sterile petri dishes in the same incubator until imaging.

Primary hippocampal neuron dissociation and culture

Neurons were dissociated from extracted hippocampi of day 19 embryos of Sprague Dawley rats, as previously described (Ortiz *et al.*, 2021). Briefly, cells were dissociated for 15 minutes from the excised hippocampus using a trypsinated buffer. The neurons were then plated onto 25 mm coverslips and incubated at 37 °C in MEM medium with 5% FBS, 2% B-27, 2% 1M dextrose and 1% GlutaMax. One day after dissociation, half the medium was exchanged for neurobasal medium with 2% B-27 supplement and 1% GlutaMax, to prevent proliferation of glial cells. Coverslips were then imaged using SRS microscopy two weeks post dissociation.

SRS microscope optical setup

The optical setup has been described previously (Kim et al., 2021) and is shown in Figure S10. Briefly, the synchronized dual output from a commercial femtosecond laser/OPO system is used for the pump and Stokes beams. The Stokes beam wavelength is fixed at 1040 nm, while the pump wavelength is tunable. For imaging at CH and CD peaks, the pump was tuned to 796 nm and 850 nm nm respectively. The Stokes beam was intensity modulated at 10.28 MHz using a quarter waveplate (Thorlabs), an electro-optic modulator (Thorlabs) driven by a function generator (Hewlett Packard) with a resonant power amplifier (Minicircuits), and a polarizer (Thorlabs). The pump and Stokes were combined colinearly using a 1000 nm short pass dichroic mirror. Temporal coincidence of the pulses was controlled using an optical delay line (Newport). A commercial scanning microscope (FV1200, Olympus) was used to raster scan the beams across the sample. A 60x 1.2 NA water immersion objective (Olympus) was used for excitation, and a 1.4 NA oil immersion condenser (Thorlabs) was used to collect the transmitted light. The Stokes beam was filtered by a 1000 nm shortpass filter, before the pump was detected on a photodiode (Hamatasu) reverse biased at 61.425 V. The stimulated Raman loss signal was extracted from the photodiode signal using a lock-in amplifier (HF2LI, Zurich Instruments).

For the images shown and transportation dynamic measurements, the setup was operated in “femtosecond” mode, where the pulses are kept temporally short to maximize peak power and thus, SRS signal. The trade off here is the relatively low spectral resolution. That is, all of the Raman transitions within the bandwidth of the lasers (~150-200 $cm^{-1}$) will be driven simultaneously. This is an acceptable trade-off in this use when just a single isolated peak (here, the CD peak) is sought. This low spectral resolution can be confounding when trying to discern multiple chemical signatures. However, for these experiments where only the dosed deuterated neurotransmitters will contribute Raman signal in the cell-silent region, The driving of all Raman modes (see Figure S2) improves sensitivity. For the cases where higher spectral resolution was desired (e.g. the acquired solution and cellular spectra shown in Figure S2), high density glass rods are used to temporally stretch and spectrally chirp the pulses so more specific Raman modes can be driven (within uncertainty of ~20 $cm^{-1}$), revealing identifying spectral characteristics of the shown molecules. Further discussion of this so-called spectral focusing can be found in Manifold et al., 2022.

Deuterated neurotransmitter imaging

PC12 cells or primary neurons cultured on coverslips were incubated for one hour at 37 °C with a final media concentration of 50 μM DA-$d_4$ or GABA-$d_6$, respectively. After one hour the coverslips were washed three times with DPBS, before being placed on a 25 mm diameter, low-profile 600 μm culture dish (Grace coverwells), filled with DPBS. The edges of the culture dish contain an adhesive which keeps the coverslips in place. PC12 cells were then imaged in the CH and CD regions using the SRS microscope. The pump and Stokes power were both set to 20 mW at focus. Images were acquired at a size of 512 x 512 pixels, using a 10 μs pixel dwell time and a lock-in constant of 3 μs (i.e. each field of view is acquired in 2.6 seconds). Neurons were imaged using a pump and Stokes power of 15 mW at focus, but otherwise identical imaging conditions. For the negative control samples for both PC12 cells and primary neurons, the media was exchanged for fresh media not containing the deuterated compound. The cells were then left to rest for one hour before being washed three times with DPBS and prepared identically to the cells incubated with the deuterated isotopologues, and imaged using the same conditions. For the post stimulation condition PC12 cells and primary neurons were incubated with 50 μM DA-$d_4$ or GABA-$d_6$, respectively. After a one hour incubation, the media was exchanged for media containing 50 mM potassium. After two minutes, the coverslips were washed twice with DPBS, and prepared and imaged analogously to the previously described conditions.

PC12 cells which were cultured on the flow cell devices were incubated with 50 μM DA-d,4 by first emptying the device of media through the open side and then introducing new media with the deuterated compound. After one hour the device was then washed twice with DPBS, before being filled with fresh media. In contrast to the PC12 cells cultured on coverslips, those cultured on the device were imaged in media not DPBS, in order to enable stimulated release of the DA-$d_4$ while on the microscope. The devices were then imaged with the same settings as those used for the cells cultured on coverslips. One CH image was taken (2930 $cm^{-1}$), then a pre-stimulation CD image, then another CD image (2150 $cm^{-1}$) following an influx of media not containing supplemental potassium, and finally a post-stimulation image 2 minutes after an influx of 50 mM potassium. This set of images was also repeated for PC12 cells cultured on devices which were not incubated with DA-$d_4$.

TPEF of FFN102 imaging

For imaging of FFN fluorescence in PC12 cells, the cells were first incubated to a final media concentration of 10 μM FFN102 for one hour at 37 °C. After one hour, samples were washed three times with DPBS. For imaging, the pump wavelength on the SRS microscope was tuned to 800 nm for two-photon excitation of FFN102, with the power set to 20 mW at focus. Images were acquired in the epi-direction at a size of 512 x 512 pixels, using a 10 μs pixel dwell time and a nondescanning PMT available on the Olympus microscope. For post potassium stimulation measurements, FFN102 was either imaged in cells 2 minutes after exposure to potassium in adjacent wells on the well plate that had not been previously imaged, or for measurements on the same cells, 2 minutes after exposure to potassium in a well that had been previously imaged.

Image processing

All statistics reported are calculated over the intensity distribution of the average signal in individual cells. For calculations involving SRS images, all images are pseudo-flatfield corrected using a replica of the respective image, convolved with Gaussian kernel of radius 200 pixels, as the flatfield. The CH channel images are then used for thresholding and segmentation. First a Gaussian blur of radius 5 pixels is applied, followed by a contrast limited adaptive histogram equalization with radius 15. Otsu's method is then applied to produce a binary thresholded image. A hierarchical contour tracing algorithm is then used to determine the outlines of the thresholded image. Internal contours, arising from organelles or low-intensity regions within a cell, are discarded, leaving the highest level outlines of the cell. Any contours with a total length of less than 100 pixels are also discarded. Each of the remaining contours is used to determine a mask over which the signal inside the contour is averaged in the corresponding CD channel. An example of this pipeline is shown in Figure S11.

Fluorescence images of PC12 cells incubated with FFN102 are processed similarly; however, as the signal is strong enough to see cell boundaries in the fluorescence channel, no additional channel is needed to serve as the mask. Instead, after flatfield correction, a second copy of each image is used for masking, and undergoes the same processing as the SRS images. The discovered contours are then used as masks on the unprocessed copy.

The images displayed in the figures have the black and white values set to reflect the dynamic range of the acquired fields of view within each experiment. This is done on a per channel basis (SRS-CH, SRS-CD, Fluorescence) so features can be readily discerned without over or under saturation. For each figure and experiment, each channel's black and white values are held constant so visual differences between different conditions are a reflection of relative molecular concentration (in SRS images) or relative fluorescence intensity.

Statistical analysis

Statistical significance, as reported in figures 3 and 7, was tested using a Welch's T-test assuming unequal variance. Effect sizes are displayed as Glass' Δ.

Supplementary Materials for

# Imaging neurotransmitter transport in live cells with stimulated Raman scattering microscopy

Bryce Manifold†, Gabriel F. Dorlhiac†, Markita Landry*, Aaron Streets*

†Co-first Authorship
*Correspondence: landry@berkeley.edu, astreets@berkeley.edu

**This file includes:**
Figs. S1 to S11

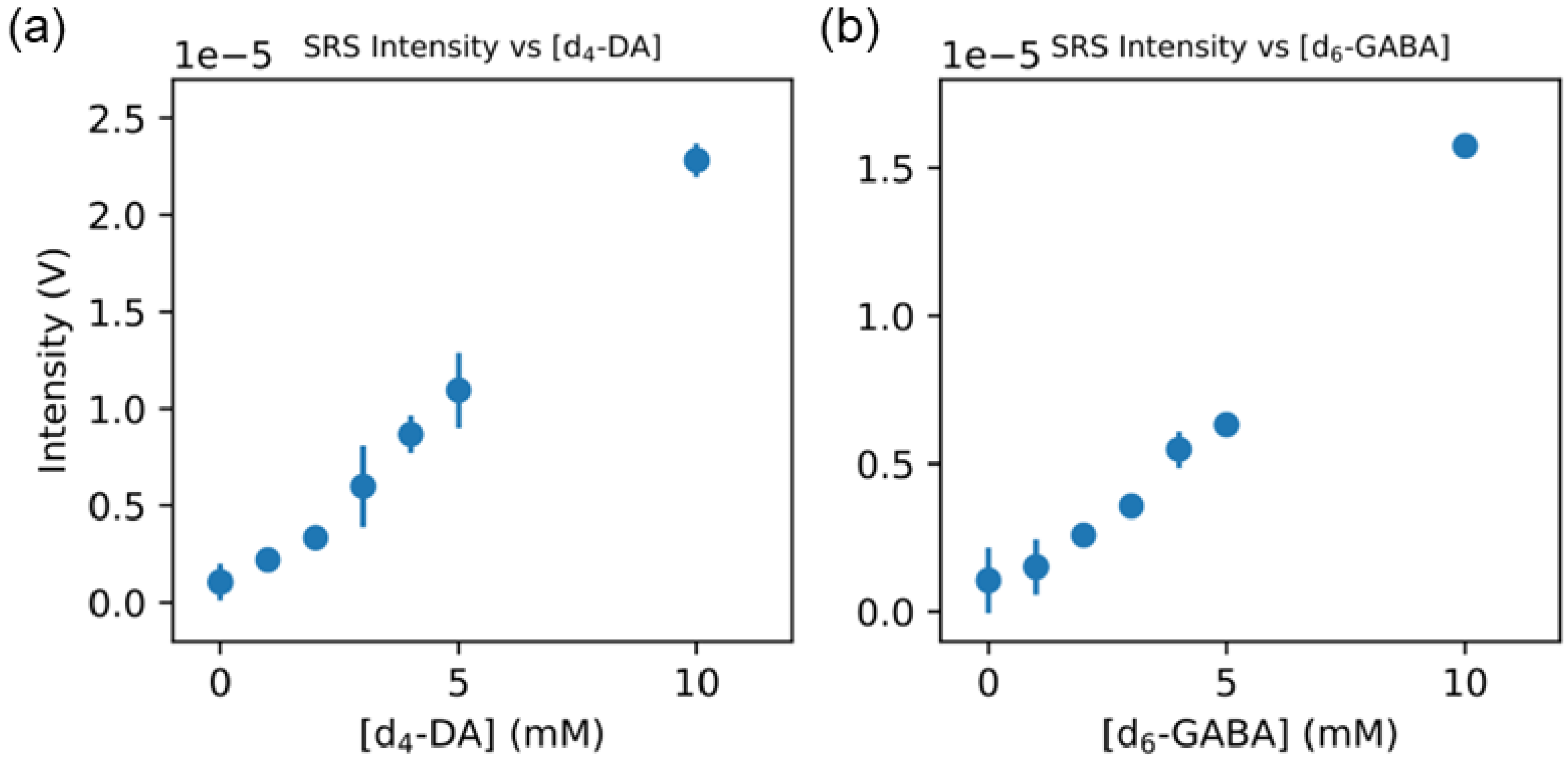

**Fig. S1. Concentration calibration curves of deuterated neurotransmitters at 2150 cm$^{-1}$.**

(a) SRS signal vs concentration of DA-$d_4$. Showing linearity down to 2 mM, and signal above background at 1 mM. (b) SRS signal vs concentration of GABA-$d_6$ showing linearity down to 2 mM and signal not significantly above background at 1 mM.

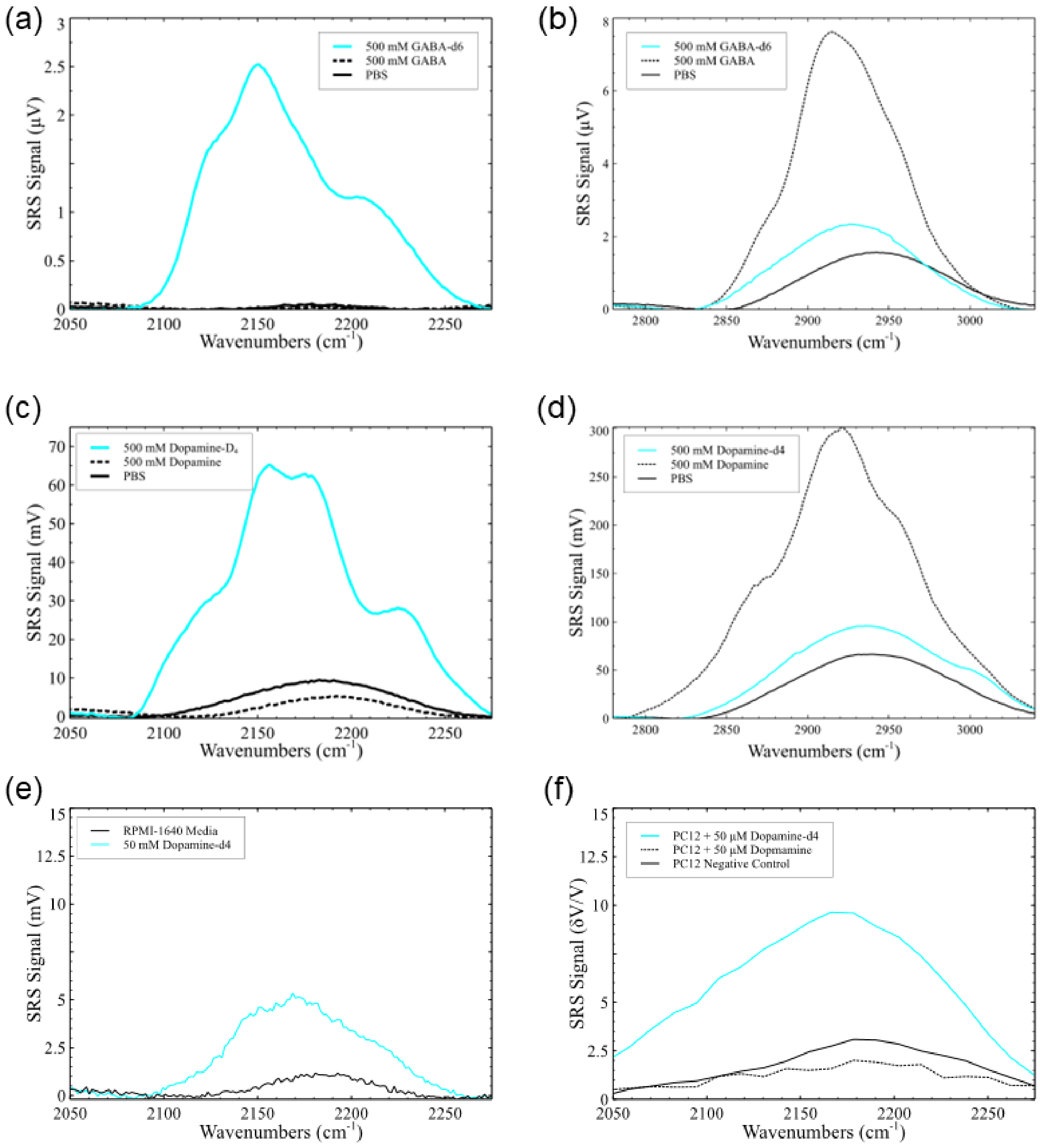

**Figure S2. SRS spectra of neurotransmitter solutions and dosed cells.**

(a) CD spectra of GABA-$d_6$, GABA, and Phosphate buffered saline (PBS). (b) CH spectra of GABA-$d_6$, GABA, and PBS. (c) CD spectra of DA-$d_4$, DA, and PBS. (d) CH spectra of DA-$d_4$, DA, and PBS. (e) CD spectra of the neuron growth medium (RPMI-1640) and the same medium containing 50 mM DA-$d_4$. (f) CD spectra of PC12 cells dosed with DA-$d_4$, DA, or equivalent PBS in medium.

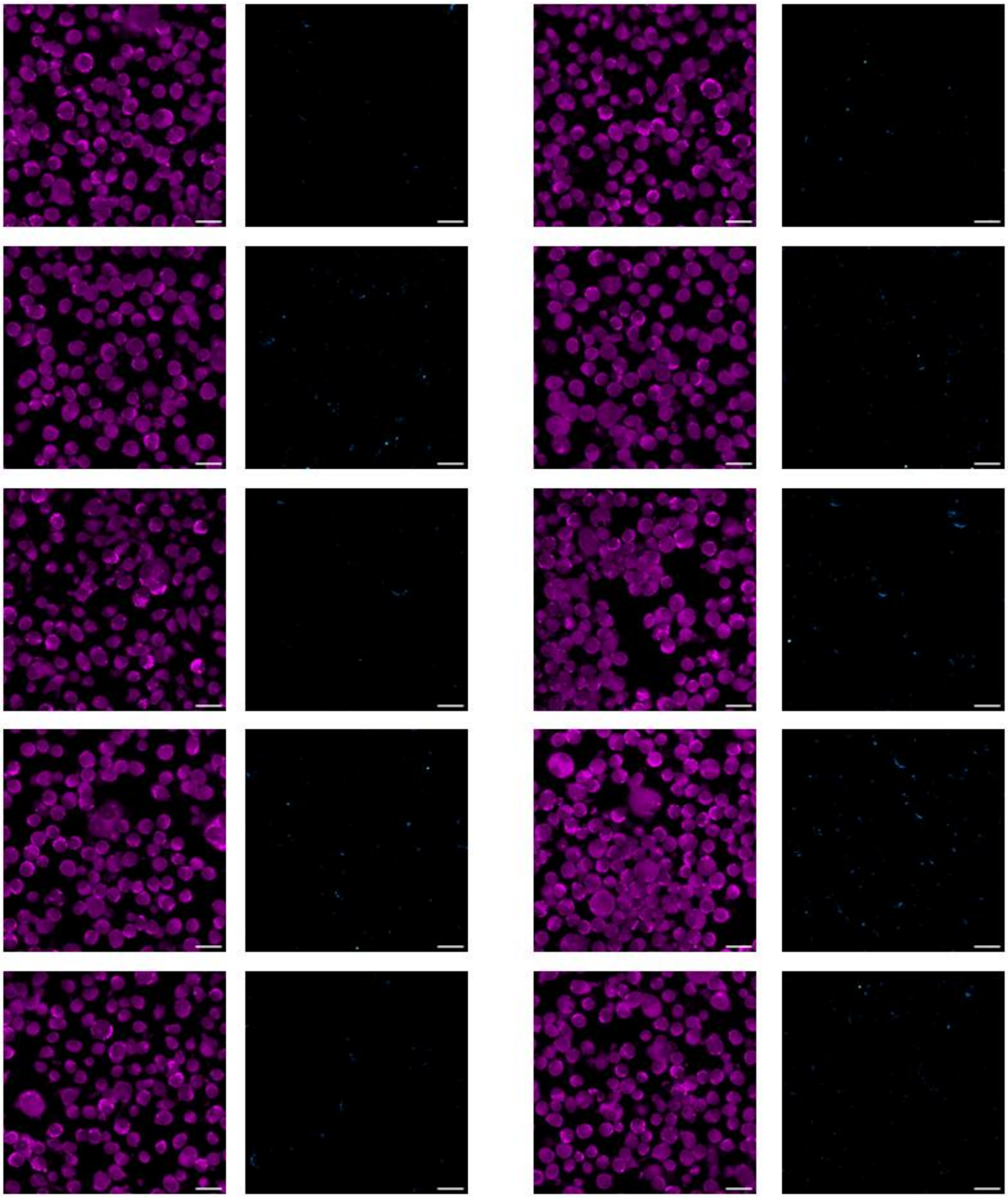

**Figure S3. Representative SRS images of PC12 cells incubated with only growth medium.**

The first and second columns, and third and fourth columns are paired images acquired of the same field of view in different SRS channels. In each pair, the left image displays

CH SRS images (magenta) and the right image displays the same field of view in CD (cyan). Scale bar = 25 μM.

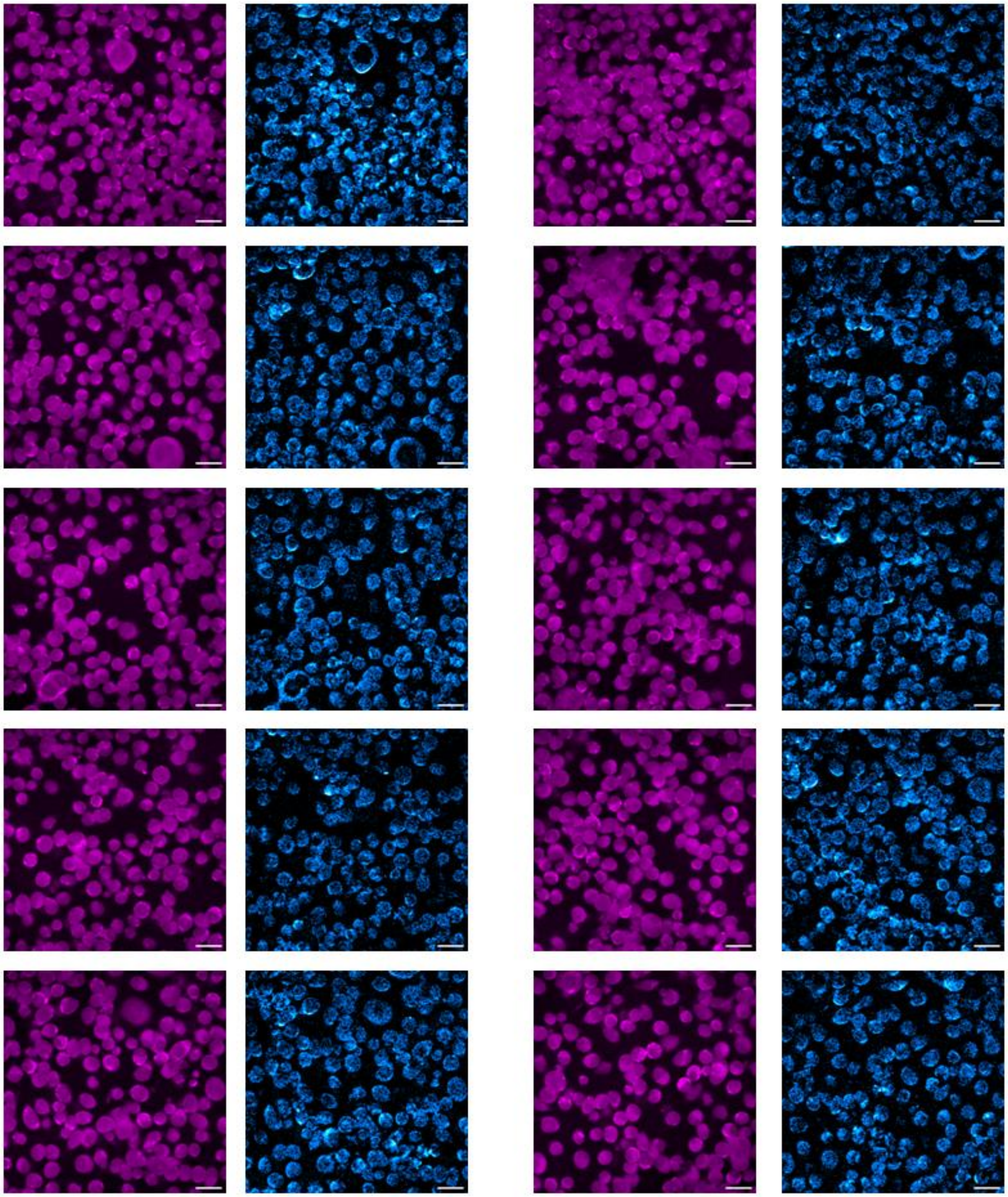

**Figure S4. Representative SRS images of PC12 cells incubated with 50μm DA-$d_4$.**

The first and second columns, and third and fourth columns are paired images acquired of the same field of view in different SRS channels. In each pair, the left image displays

CH SRS images (magenta) and the right image displays the same field of view in CD (cyan). Scale bar = 25 μM.

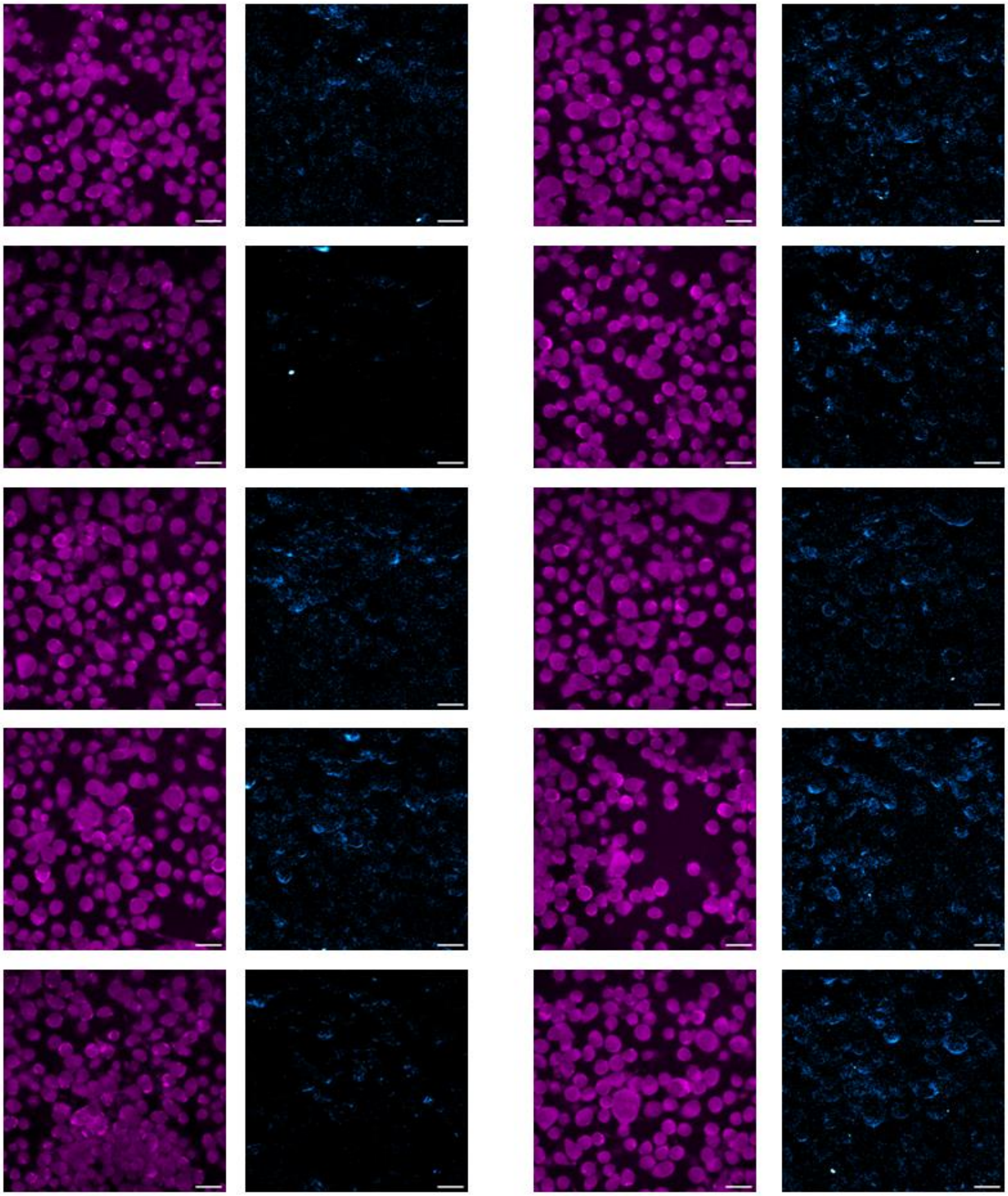

**Figure S5. Representative SRS images of PC12 cells incubated with 50μm DA-$d_4$ followed by stimulation with 50 mM potassium.**

The first and second columns, and third and fourth columns are paired images acquired of the same field of view in different SRS channels. In each pair, the left image displays CH SRS images (magenta) and the right image displays the same field of view in CD (cyan). Scale bar = 25 μM.

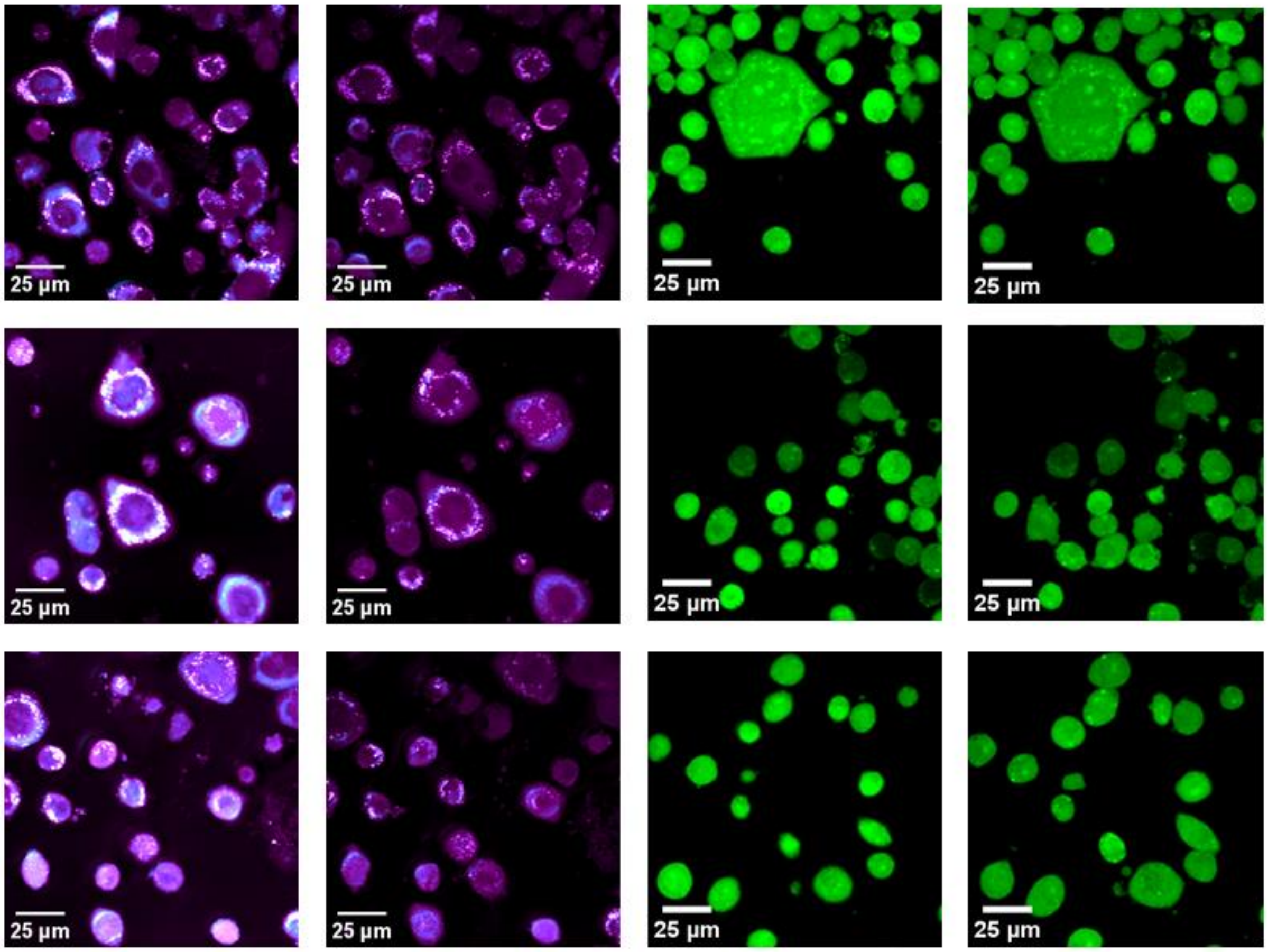

**Figure S6. PC12 Cells before and after potassium simulation as imaged on the custom microfluidic chip.**
Left two columns show example fields of view of PC12 cells dosed with 50μm before and after potassium stimulation on the chip. Here the images are a composite of CH (magenta) and CD (cyan) SRS signal. The measured decrease in SRS intensity at the CD peak is seen visually. The right two columns show the same experiment, but using two-photon fluorescence of the FFN102.

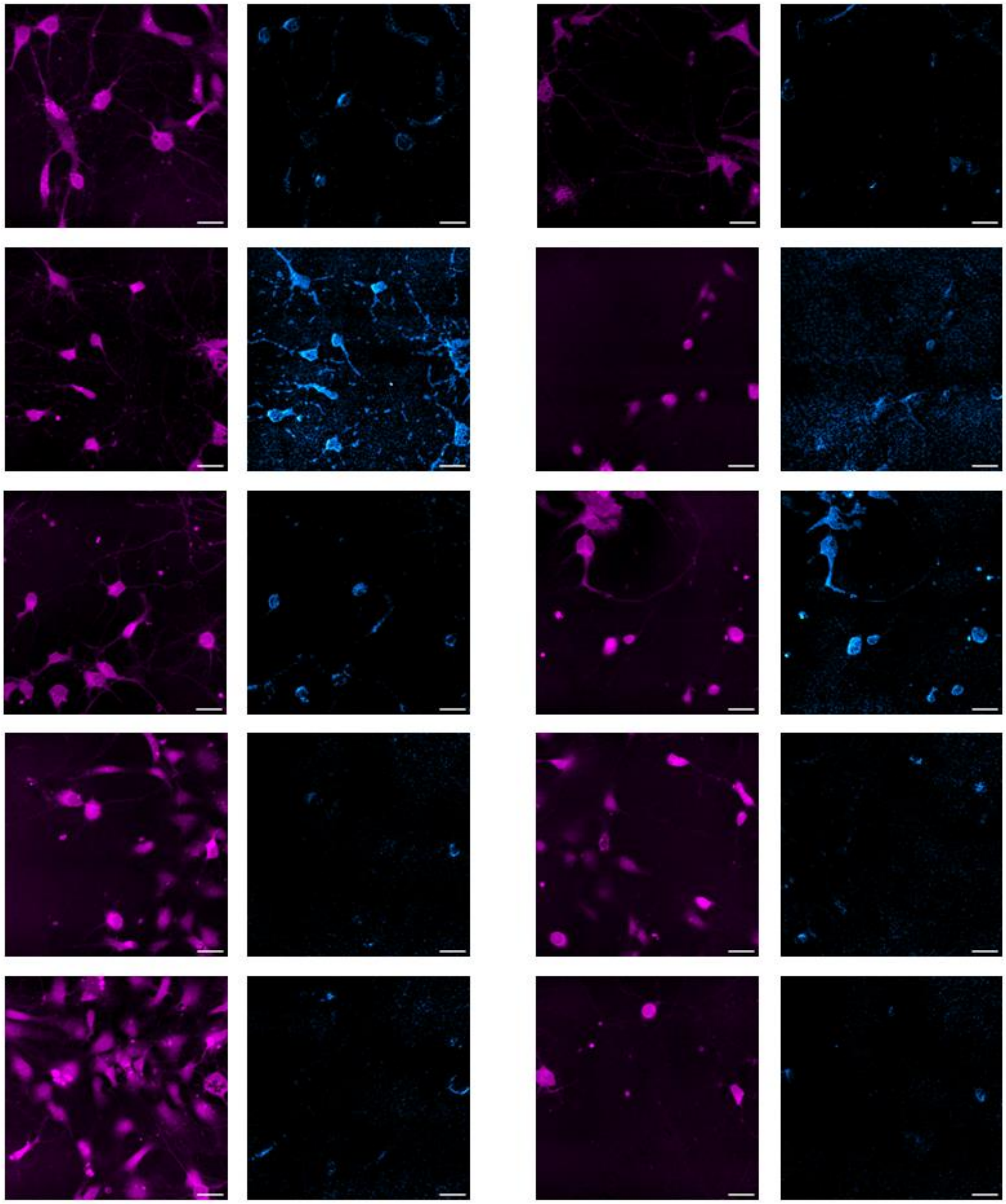

**Figure S7. Representative SRS images of Hippocampal neurons incubated with 50µm GABA-$d_6$.**

The first and second columns, and third and fourth columns are paired images acquired of the same field of view in different SRS channels. In each pair, the left image displays CH SRS images (magenta) and the right image displays the same field of view in CD (cyan). Scale bar = 25 µM.

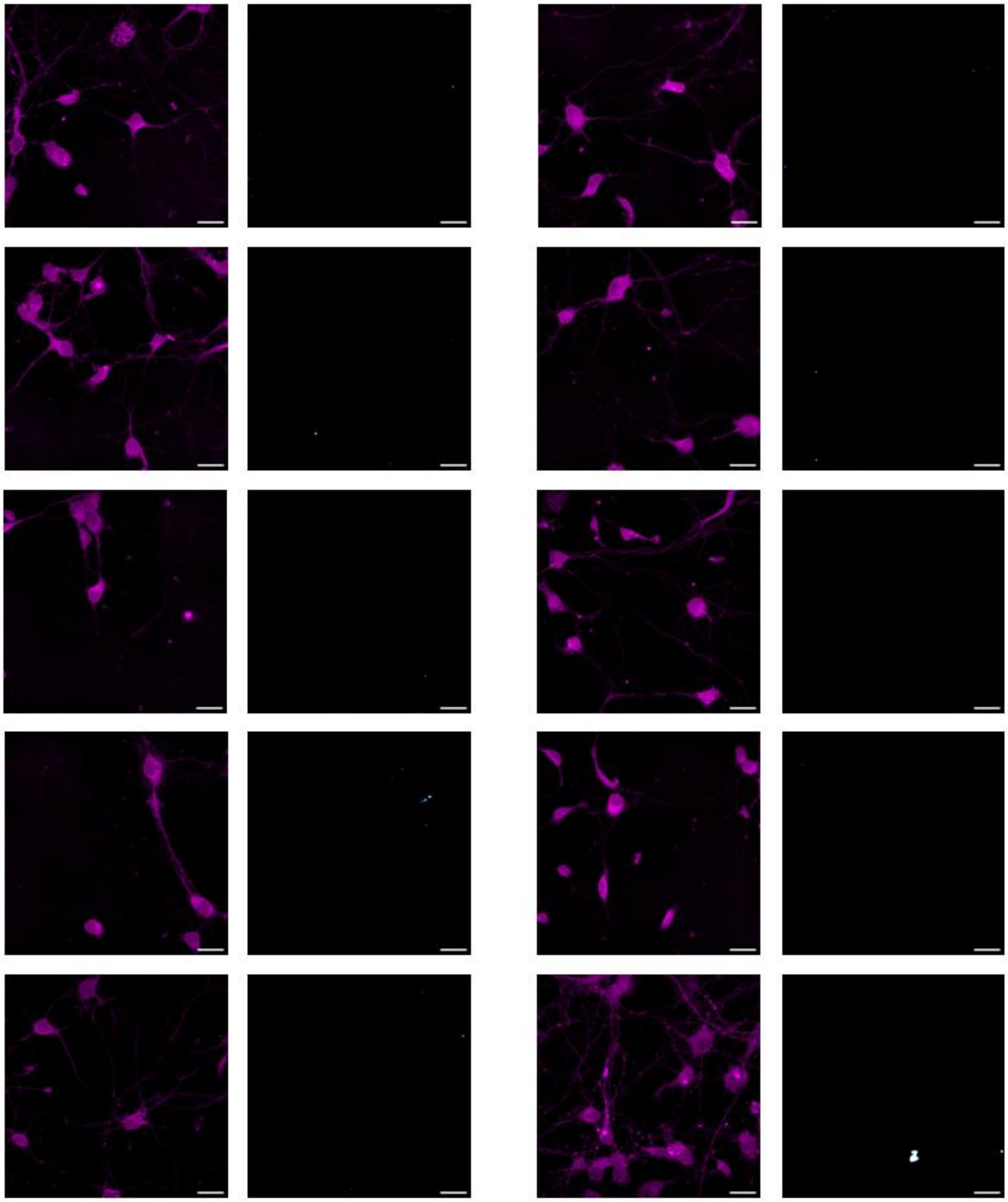

**Figure S8. Representative SRS images of Hippocampal neurons incubated with 50μm GABA-$d_6$ followed by stimulation with 50 mM potassium.**

The first and second columns, and third and fourth columns are paired images acquired of the same field of view in different SRS channels. In each pair, the left image displays CH SRS images (magenta) and the right image displays the same field of view in CD (cyan). Scale bar = 25 μM.

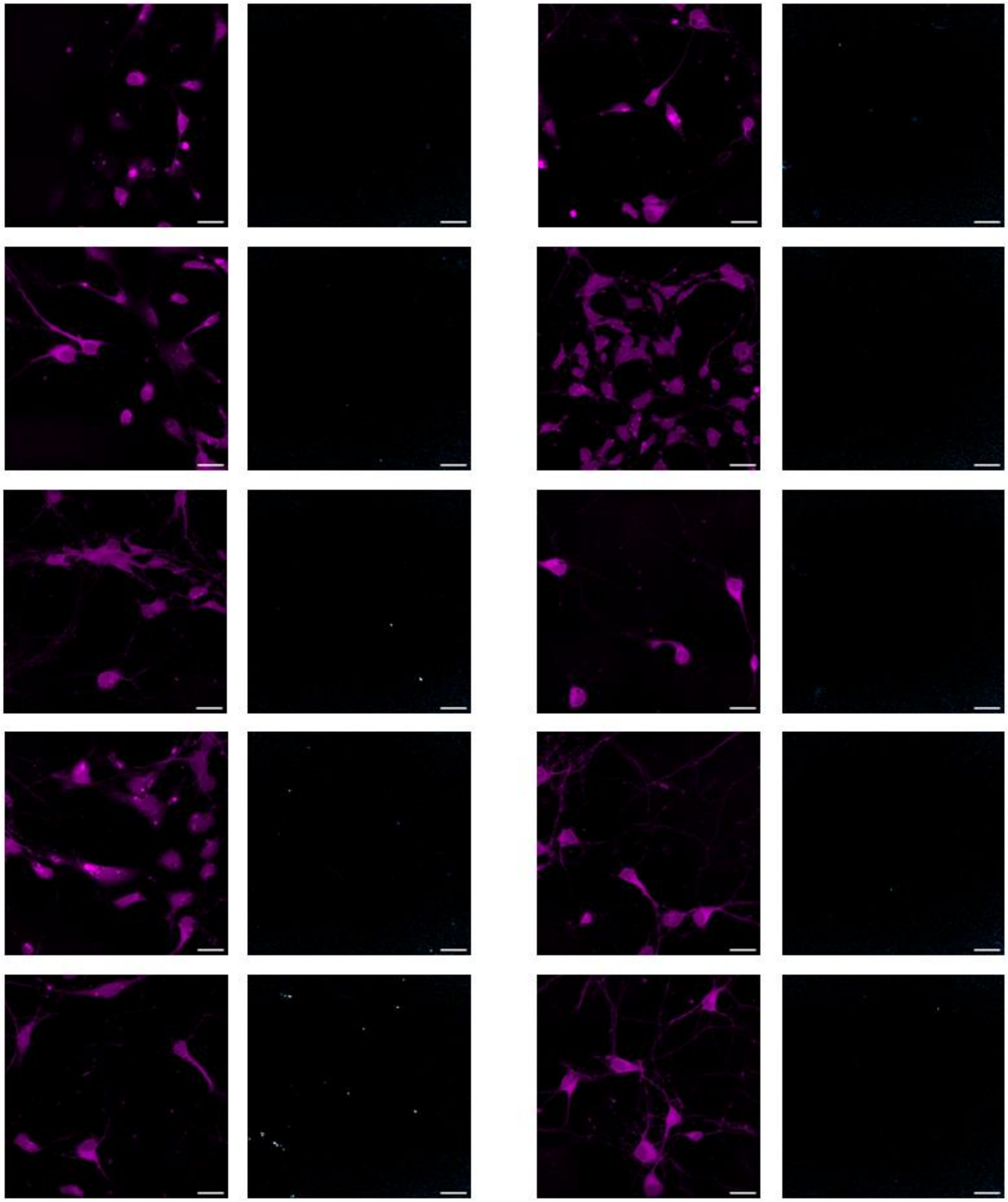

**Figure S9. Representative SRS images of Hippocampal neurons incubated only with growth medium.**
The first and second columns, and third and fourth columns are paired images acquired of the same field of view in different SRS channels. In each pair, the left image displays CH SRS images (magenta) and the right image displays the same field of view in CD (cyan). Scale bar = 25 μM.

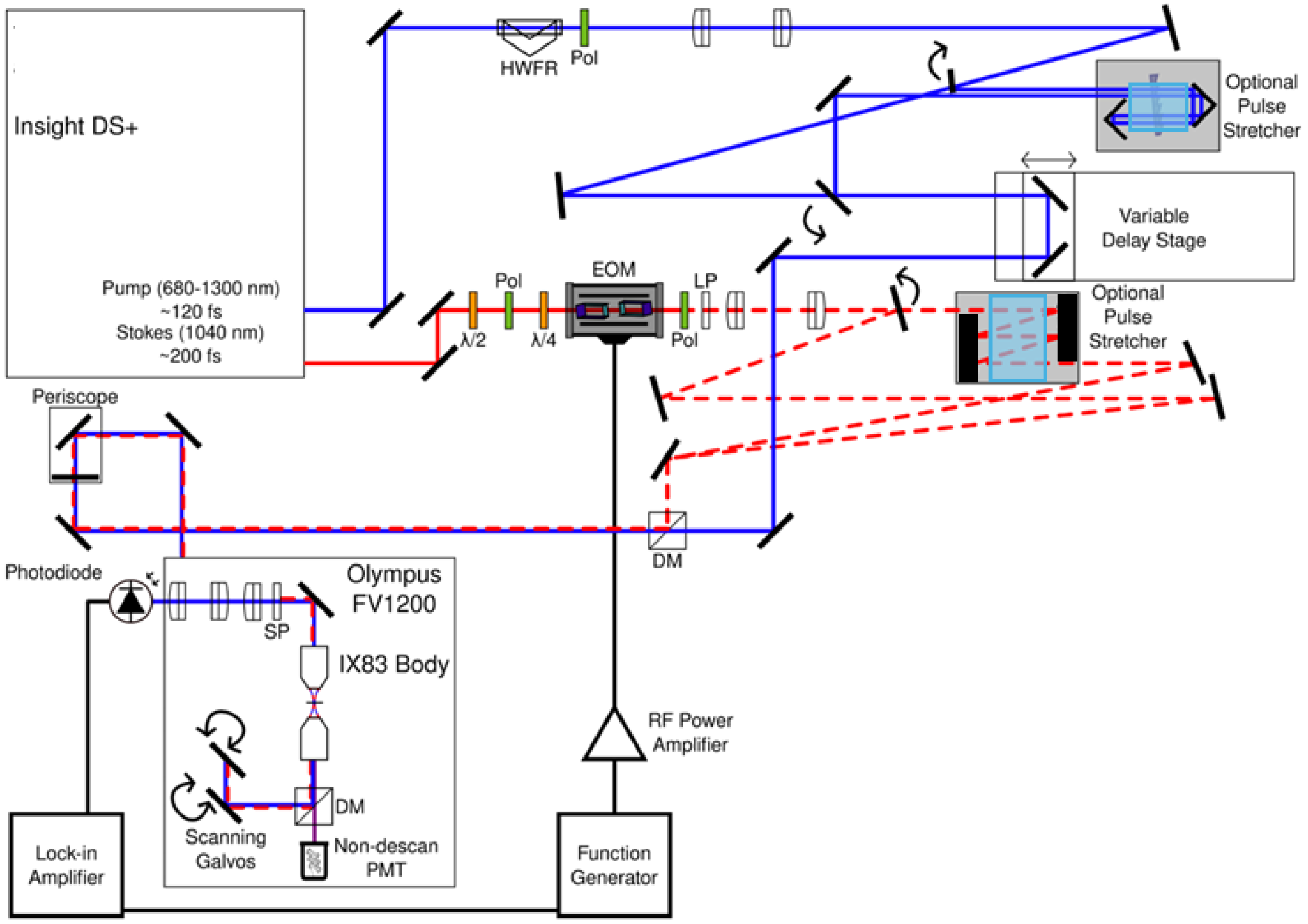

**Figure S10. Schematic of SRS microscope optical setup.**

Solid black lines represent mirrors, black lines with ridges are reflective gratings, with the translucent variant being a transmission grating. Lenses are doublets depicted as unlabeled white components outlined in black. EOM: Electro-optic modulator; λ/2: half-wave plate; Pol: polarizer; λ/4: quarter-wave plate; LP: long-pass filter; DM: dichroic mirror; SP: short-pass filter.

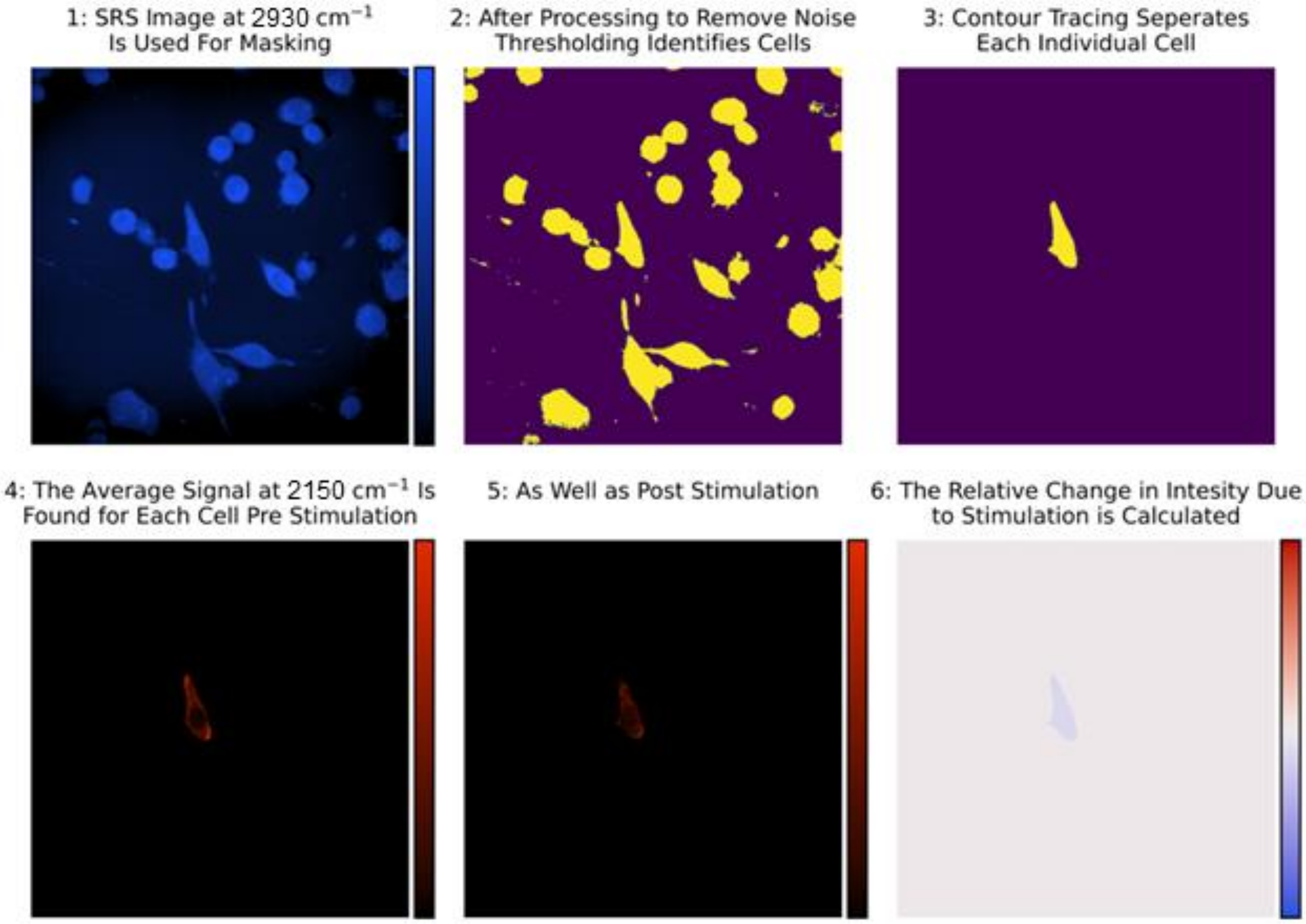

**Figure S11. Visual example of image analysis pipeline.**